\documentstyle[psfig,pre,aps]{revtex}

\newcommand{\be}{\begin{equation}}
\newcommand{\ee}{\end{equation}}

\begin{document}
\draft
\twocolumn[\hsize\textwidth\columnwidth\hsize\csname
@twocolumnfalse\endcsname
\title{Dynamics of spherical particles on a surface:\\ 
About collision induced sliding and other effects}
\author{Ljubinko Kondic~\cite{email}}
\address{Departments of Mathematics and Physics,
Duke University, Durham, NC 27708}
\date{\today}
\maketitle
\begin{abstract}
We present a model for the motion of hard spherical particles on a two dimensional
surface. The model includes both the interaction between the particles via
collisions, as well as the interaction of the particles with the substrate.
We analyze in details the effects of sliding and rolling friction, that are 
usually overlooked.  It is found that the properties of this particulate system
are influenced significantly by the substrate-particle interactions. 
In particular, sliding of the particles relative to the substrate {\it after}
a collision leads to considerable energy loss for common experimental conditions.  
The presented results provide a basis, that can be used to realistically model
the dynamical properties of the system, and provide further insight into
density fluctuations and related phenomena of clustering, structure formations, 
and inelastic collapse.   
\end{abstract}
]

\section{Introduction}

In this paper we address the problem of the motion of a set of hard 
spherical particles on an inclined, in general dynamic surface.
While there have been substantial efforts to understand in more details 
the problem of the nature of interaction of a single particle with the 
substrate~\cite{rabinowicz,baumberger,reynolds,tabor,green,fuller,brilliantov2,domenech,ehrlich,gersten},
these efforts have not being extended to the multiparticle situation. 
On the other hand, there has been recently a lot of interest
in one~\cite{bernu,mcnamara1,goldhirsch2,grossman1,constantin,kadanoff} 
or two~\cite{mcnamara2,mcnamara3,goldhirsch1,ben-naim,golob,behringer,ben,we} 
dimensional granular systems.  These systems are of considerable importance, 
since they provide useful insight into more complicated systems arising in 
industrial applications, and also because of many fascinating
effects that occur in simple experimental settings
and theoretical models. Theoretical and computational efforts have lead to results including
density fluctuations, clustering, and inelastic
collapse~\cite{mcnamara1,goldhirsch2,grossman1,kadanoff,mcnamara2,mcnamara3,goldhirsch1,ben-naim}.  
Further, a system of hard particles energized by either an oscillating side wall, 
or by oscillating surface itself, has been explored recently~\cite{grossman1,kadanoff,ben-naim}. 
This system, due to its similarity to one or two dimensional gas, appears
to be a good candidate for modeling using continuous hydrodynamic 
approach~\cite{grossman1,kadanoff,ben-naim}.  

It is of great interest to connect theoretical and computational
results with experimental ones.  Very recently, it has been
observed experimentally that many complex phenomena
occur in the seemingly simple system of hard particles rolling and/or sliding 
on a substrate.  In particular, clustering~\cite{golob,behringer}, and 
friction-based segregation~\cite{behringer} have been observed. 
While some of the experimental results (e.g. clustering) could
be related to the theoretical results~\cite{ben-naim}, there are still
considerable discrepancies.  Theoretically, it has been found
that the coefficient of restitution, measuring the elasticity of 
particle-particle collisions, is the important parameter of the problem, 
governing the dynamical properties of the system. 
While the coefficient of restitution is definitely an important quantity,
a realistic description of an experimental system cannot be based just on
this simple parameter.  As pointed out in~\cite{golob}, the rotational motion of 
the particles and the interaction with the substrate introduce an additional
set of parameters (e.g. the coefficients of rolling and sliding friction),
that have not been included in the theoretical descriptions of the system.

Our goal is to bridge this gap between experiment and theory, and
formulate a model that includes both particle-particle
and particle-substrate interactions, allowing for a comparison
between experimental and theoretical results.
Specifically, we address the phenomena of rolling friction and sliding, 
that lead to the loss of mechanical energy and of linear and 
angular momentum of the particles.  In order to provide a better understanding 
of the importance of various particle and substrate properties that 
define the system (e.g. rolling friction, sliding, the inertial properties
of the particles), we concentrate part of the discussion on monodisperse, 
hard (steel), perfectly spherical solid particles, moving on a hard (aluminum, copper) 
substrate~\cite{golob,behringer,ben}.  However, through most of the presentation, 
the discussion is kept as general as possible,
and could be applied to many other physical systems.  Specifically, the 
extension of the model to more complicated systems and geometries is of
importance, since most of the granular experiments involve some kind of 
interaction of the particles with (static or dynamic) walls.  In particular, 
the discussion presented here is relevant to wall shearing experiments, where
particle-wall interactions are of major importance in determining the 
properties of the system (see~\cite{behringer_pt} and the references therein).

In section~\ref{sec:forces} we explore all of the forces that
act on the particle ensemble on a moving, inclined substrate.  
First we explore particle-particle interactions,
and formulate a model incorporating the fact that the particles roll on
a surface, and have their rotational degrees of freedom 
considerably modified, compared to ``free'' particles.  
Further, we include the interaction of the particles with the
substrate, paying attention to the problem of rolling
friction and sliding.  The analysis is extended to the situation
where the substrate itself is moving with the prescribed velocity
and acceleration.  Because of the complexity of the interactions that
the particles experience, we first consider the
problem of particles moving without sliding, and include the 
sliding at the end of the section.  In section~\ref{sec:move}, 
we give the equations of motion for a particle that experiences collisions 
with other particles, as well as interaction with the substrate.  
In Section~\ref{sec:discussion} we apply these equations of
motion to the simple case of particles moving in one direction only.
It is found that many interesting effects could be observed in this
simple geometry.  In particular, we explore the effect of sliding
both during and after the collisions, and give estimates for the 
experimental conditions that lead to sliding.  Finally, we give the
results for the time the particles slide after a collision, 
as well as for the sliding distance, and for the loss of the translational 
kinetic energy and linear momentum of the particles.

\section{Forces on particles}\label{sec:forces}

Particles moving on an inclined hard surface experience three
kind of forces:
\begin{itemize}
\item Body forces (gravity);
\item Forces due to collisions with other particles and walls;
\item Forces due to interaction with the substrate.
\end{itemize}
In what follows we analyze each of these forces, with
emphasis on understanding the interaction between substrate and
the particles.  While the analysis is kept as general as possible, some
approximations appropriate to the problem in question
are utilized, in order to keep the discussion tractable.  
In particular, the coefficient of rolling friction is assumed
much smaller then (static and kinematic) coefficients of sliding 
friction.  Further, in this section the particles are confined to
move on the substrate without jumping; the experimental conditions under 
which this extra degree of freedom is introduced are discussed in
section~\ref{sec:discussion}.  Throughout most of this section it is 
assumed that the particles are moving on the substrate without sliding; in 
section~\ref{sec:slip} we explain the conditions for sliding to
occur. 

In order to formulate a model that can be used for efficient molecular
dynamics simulations, we choose rather simple models for the interactions
between the particles and between the particles and the substrate.  In
modeling collisions between particles,, 
we neglect static friction, as it is often
done~\cite{haff_werner,campbell,hermann,thompson,luding,ristow}.  On
the other hand, the static friction between the particles and the
substrate is of major importance, since it leads to rolling particle 
motion; consequently, it is included in the model.  

\subsection{Body forces}

Here we consider only gravitational force that acts on the center
of mass of the particles.  It is assumed that there are no other
(e.g. electrostatic) long range forces.  In the coordinate frame
that is used throughout (see Fig. 1), the acceleration
of a particle due to gravity, $g$, is
\begin{equation}
{\bf a}_G = - {\bf \hat j}\, g \sin(\theta)\ ,
\label{eq:grav}
\end{equation}
where $\theta$ is the inclination angle.
\begin{figure}[htb]
\centerline{\psfig{figure=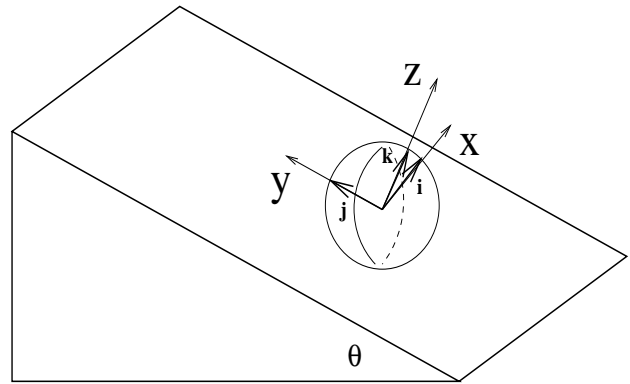,height=2.0in,width=3.25in}}
\vspace{0.1in}
\caption{The coordinate frame used in the paper.}
\end{figure}

\subsection{Collision forces}
\label{sec:coll}

There are many approaches to modeling collision 
interactions between particles (see, e.g.~\cite{hermann,haff} and references therein).  
We note that rather complex models have been 
developed~\cite{cundall,walton_braun,walton,goddard,taguchi,radjai1,radjai2,radjai3,brilliantov1,poschel}, 
but choose to present a rather simple one, which,
while necessary incomplete, still models realistically collision
between particles moving with moderate speeds. In the context of the particles moving on a 
substrate, it is important to realize that, even though particles are 
confined to move on a 2D surface, the 3D nature of the particles is of importance.  
Even if one assumes that the particles roll on a substrate without
sliding, only two components ($x$ and $y$) of their angular velocity,
${\bf \Omega}$, are determined by this constraint. The particles could 
still rotate with $\Omega^z$, which could be produced by collisions
(hereafter we use ${\bf \Omega}$ to denote the components of angular
velocity in $x-y$ plane only).  We will see that rotations of the particles 
influence the nature of their interaction, as well as the interaction
with the substrate.

{\it Normal force.}
Using a simple harmonic spring model~\cite{hermann,thompson,haff}, 
the normal force on
particle $i$, due to the collision with particle $j$, is given by
\begin{equation}
{\bf F}_N^c = \left [ k (d - r_{i,j}) -
\gamma_N \bar m ({\bf v}_{i,j} \cdot {\bf\hat n})\right ] 
{\bf\hat n}\ ,
\label{eq:fn}
\end{equation}
where $k$ is a force constant, $r_{i,j} = |{\bf r}_{i,j}|$,
${\bf r}_{i,j} = {\bf r}_i - {\bf r}_j$, 
${\bf \hat n} = {{\bf r}_{i,j}/ r_{i,j}}$, 
${\bf v}_{i,j} = {\bf v}_i - {\bf v}_j$,
$\bar m$ is the reduced mass, and
$d = R_i + R_j$, where $R_i$ and $R_j$ are the radii of the particles 
$i$ and $j$, respectively.  In this paper, we assume monodisperse 
particles, so that $d$ is equal to the diameter of a particle. 
The energy loss due to inelasticity of the collision is included by the damping constant,
$\gamma_N$.  The damping is assumed to be proportional to 
the relative velocity of the particles in the normal direction, ${\bf \hat n}$.
While we use this simple linear model, the parameters $k$ and $\gamma_N$ are
connected with material properties of the particles using nonlinear (Hertzian)
model (see Appendix~\ref{sec:nonlinear}).  We note that $\gamma_N$ is connected with 
the coefficient of restitution, $e_n$, by $e_n = \exp({-\gamma_n t_{col}/2})$.  
Here $t_{col}$ is the collision time and is approximately given by 
$t_{col}\approx\pi \sqrt{m/(2k)}$ (see Appendix~\ref{sec:linear}).  
While more realistic nonlinear models lead to the velocity dependent coefficient of 
restitution~\cite{walton_braun,brilliantov1,poschel}, the linear
model is satisfactory for our purposes, since we are interested 
in the collisions characterized by moderate impact velocities.
In the case of particle-wall collisions, $d\rightarrow d/2$,
${\bar m} \rightarrow m$, and ${\bf r}_{i,j} \rightarrow 
{\bf r}_{wall} - {\bf r}_i$, where ${\bf r}_{wall}$ is the coordinate of 
the contact point at the wall.

{\it Tangential force in $x-y$ plane.}
The motion of the particles in the tangential direction (perpendicular to 
the normal direction, in $x-y$ plane), leads to a tangential (shear) force.
This force opposes the motion of the interacting particles in the 
tangential direction, so that it acts in the direction that is 
opposite to the relative tangential velocity, $v_{rel}^t$, of the 
point of contact of the particles.  Both translational motion of 
the center of mass, and rotations of the particles with component
of angular velocity in the ${\bf \hat k}$ direction contribute to $v_{rel}^t$, thus
\begin{equation}
v_{rel}^t = {\bf v}_{i,j}\cdot {\bf \hat s} + 
{d\over 2}\; (\Omega_i^z + \Omega_j^z)\ ,  
\label{eq:vrel}
\end{equation}
where ${\bf \hat s} = ({\bf \hat n}\cdot {\bf \hat j},
-{\bf \hat n}\cdot {\bf \hat i})$.  We model this 
force (on the particle $i$) by~\cite{hermann,thompson}
\begin{equation}
{\bf F}_S^c =  sign(-v_{rel}^t) min \left ( \gamma_S \bar m |v_{rel}^t|, \nu_k
|{\bf F}_N^c |\right ) {\bf\hat s}\ .
\label{eq:shear}
\end{equation}
Here the Coulomb proportionality between normal and shear (tangential)
stresses requires that the shear force, $|{\bf F}_S^c|$, is limited
by the product of the coefficient of kinetic friction between the
particles, $\nu_k$, and the normal force, $|{\bf F}_N^c|$.  
The damping coefficient in the tangential (shear) direction,
$\gamma_S$, is usually chosen as $\gamma_S = {\gamma_N/2}$, so that the coefficients
of restitution in the normal and shear directions are identical~\cite{thompson}.
An alternative method, where one models shear force by introducing
a ``spring'' in the tangential direction, and calculates
the force as being proportional to the extension of this spring, 
has been used as well (see, e.g.~\cite{haff,cundall}). We neglect static
friction~\cite{haff_werner,campbell,hermann,thompson,luding,ristow,brilliantov1}.

The torque on the particle $i$, due to the force ${\bf F}_S^c$, is
${\bf T}_i = {\bf x}_i \times {\bf F}_S^c$, where ${\bf x}_i$ is
the vector from the center of the particle $i$ to the point
of contact, so ${\bf x}_i = -{d/2}\; {\bf \hat n}$.  This torque
produces the angular acceleration of the particle $i$ in the 
${\bf \hat k}$ direction, $ \dot \Omega_i^z = {T_i^z/I} = 
-{d/(2 I)}\, {\bf \hat n}\times {\bf F}_S^c$, 
where $I$ is the particles' moment of inertia.
Recalling that ${\bf F}_S^c$ is defined by~Eq.(\ref{eq:shear}), one obtains
(we drop subscript $i$ hereafter, if there is no possibility for confusion)
\begin{equation}
\dot \Omega^z = - {d\over 2 I} |{\bf F}_S^c| sign(v_{rel}^t)  \ .
\nonumber
\end{equation}
By direct integration, 
this result yields $\Omega^z$. With this the relative
velocity, $v_{rel}^t$ (Eq.~(\ref{eq:vrel})), is given, and
hence we can calculate the tangential (shear) force, ${\bf F}_S^c$. 

{\it Tangential force in the ${\bf \hat k}$ direction}.
Since the particles are rolling, there is an additional
force due to the relative motion of the particles at the point of 
contact in the perpendicular, ${\bf \hat k}$, direction.  Figure 2
gives a simple example of the collision of two particles with translational
velocities in the ${\bf \hat i}$ direction only.
We model this force, ${\bf F}_R^c$, which is due to rotations of the 
particles with angular velocity, ${\bf \Omega}$,
in the same manner as the shear force, ${\bf F}_S^c$.
The force, ${\bf F}_R^c$, on the particle $i$ due to a collision with
the particle $j$, acts in the direction opposite to the ${\bf \hat k}$
component of the relative velocity of the point of contact.
Similarly to the ``usual'' shear force, we assume that 
the magnitude of ${\bf F}_R^c$ cannot be larger than
the normal force times Coulomb coefficient; thus

\be
{\bf F}_R^c =  sign(-v_{rel}^z) min(\gamma_s {\bar m} |v_{rel}^z|,
\nu_k |{\bf F}_N^c|) {\bf \hat k}\ ,
\label{eq:force_r}
\ee
and, for a general collision,
$v_{rel}^z = R[({\bf \Omega}_i+{\bf \Omega}_j)\times {\bf \hat k}]\cdot {\bf \hat n}$.
In the case of a central collision as shown in Fig. 2, $v_{rel}^z$ simplifies to 
\be
v_{rel}^z = R({\bf \Omega}_i+{\bf \Omega}_j)\cdot {\bf \hat j}\ .
\label{eq:vrelz}
\ee
\begin{figure}
\centerline{\psfig{figure=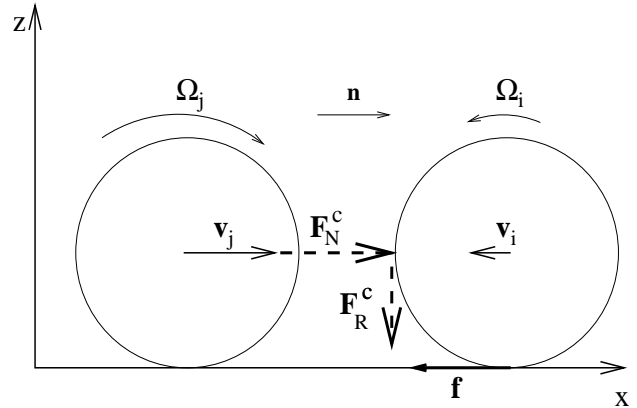,height=2.15in,width=3.25in}}
\vspace{0.1in}
\caption{The collision between two particles, with the linear velocities
in the ${\bf \hat i}$ direction only.  The direction of ${\bf F}_R^c$
follows from the assumption that $|{\bf v}_j| > |{\bf v}_i|$.
The friction force ${\bf f}$ is explained in the following
section.  (The particles $i$ and $j$ are assumed to be in contact; for clarity
reasons we show them separated). 
}
\end{figure}
The torque acting on the particle $i$ due to this force is
${\bf T} = {\bf x} \times {\bf F}_R^c$. This torque produces an angular acceleration 
${\bf \dot \Omega} = {{\bf T}/I}$.  Assuming that there is no sliding of the particles 
with respect to the substrate, we obtain the following result for the linear
acceleration of the particle $i$
\be
{\bf a}_R^c = R {\bf \dot \Omega} \times {\bf \hat k} =
{R^2 |F_R^c|\over I} \, sign(-v_{rel}^z) {\bf \hat n}\nonumber \ .
\ee
A 2D example given in Fig. 2 shows that
${\bf a}_R^c$ of both particles, $i$ and $j$, is in the direction of 
${\bf \hat n}$, or $-{\bf \hat i}$ direction.

Let us also note that the force ${\bf F}_R^c$ modifies the normal
force, ${\bf F}_N$,  with which the substrate acts on the particle.
From the balance of forces in the ${\bf \hat k}$ direction, it follows that
the normal force is given by 
\be
{\bf F}_N = mg {\bf \hat k}\cos(\theta) - 
{\bf F}_R^c \ .
\label{eq:fn_mod}
\ee
The ``jump'' condition, $mg {\bf \hat k}\cos(\theta) = {\bf F}_R^c$ is discussed
in more details in section~\ref{sec:discussion}.
Here we assume nonzero ${\bf F}_N$, and consider only the motion in $x-y$ plane.

To summarize, the collision interactions of the particle $i$ with
the particle $j$ lead to the following expression for the total 
acceleration of particle $i$ (in $x-y$ plane)
\be
{\bf a}^c = {\bf a}_N^c + {\bf a}_S^c +{\bf a}_R^c\ ,
\label{eq:coll}
\ee
where 
\begin{eqnarray}
{\bf a}_N^c =&& {1\over m} \left [ k (d - r_{i,j}) -
\gamma_N \bar m ({\bf v}_{i,j} \cdot {\bf\hat n})\right ]
{\bf\hat n} \nonumber \ , \\ 
{\bf a}_S^c =&& {1\over m}\, min \left ( \gamma_S \bar m |v_{rel}^t|, \nu_k
|{\bf F}_N^c |\right ) \, sign(-v_{rel}^t) {\bf\hat s}\nonumber \ ,\\
{\bf a}_R^c =&& {R^2\over I}\, min(\gamma_s {\bar m} |v_{rel}^z|,
\nu_k |{\bf F}_N^c|) \, sign(-v_{rel}^z) {\bf \hat n}\nonumber\ .
\end{eqnarray}
Here
${\bf a}_N^c$ is the acceleration due to the normal force given
by Eq.~(\ref{eq:fn}), ${\bf a}_S^c$ is the tangential acceleration due to the
shear force, given by Eq.~(\ref{eq:shear}),
and ${\bf a}_R^c$ is the rotational acceleration due to the 
tangential force in the ${\bf \hat k}$ direction, given by 
Eq.~(\ref{eq:force_r}).

\subsection{Interaction with the substrate}

The theory of rolling and sliding
motion of a rigid body, even on a simple horizontal 2D substrate is 
complicated.  For example, even though the question of
rolling friction was addressed long ago~\cite{reynolds}, more recent
works~\cite{tabor,green,fuller,brilliantov2,domenech,ehrlich,gersten}
show that there are still many open questions about the origins of rolling friction;
similar observation applies to sliding friction.  In order
to avoid confusion, we use term ``friction'' to refer to either static or kinematic
(sliding) friction; rolling friction is considered separately.

We approach this problem in several steps.
After the introduction of the problem, we first consider 
a particle rolling without sliding, with vanishing
coefficient of rolling friction, $\mu_r$.  Next we present the generalization, 
that allows for nonzero $\mu_r$, as well as for the possibility of sliding.
The substrate is assumed to move with its own prescribed velocity, ${\bf v}_S$, 
and acceleration, ${\bf a}_S$, which could be time dependent.  The generalization
to space dependent  ${\bf v}_S$ and ${\bf a}_S$ is straightforward, but
it is not introduced for simplicity.  Similarly, we assume that the substrate
is horizontal; the generalization to an inclined substrate is obvious.

Figure 3 shows the direction of the forces acting on a rolling particle.
The friction force ${\bf f}$, that causes the particle to roll, 
acts in such a direction  to produce the torque, ${\bf T}$, in the direction of 
the angular acceleration of the particle. 
Assuming that this friction force is applied to the instantaneous rotation
axis, it does not lead to a loss of mechanical energy, 
as pointed out in~\cite{domenech}.  If $\mu_r$ is
zero, the particle will roll forever on a horizontal surface.  

On the other hand, the rolling friction force, ${\bf f}_r$,
acts in such a way to oppose the rotations. Thus it produces the 
torque, ${\bf T}_r$, in the direction opposite to the angular velocity of
the particle.  This torque could be understood if one assumes
a small deformation of the substrate and/or particle, that modifies the 
direction of the rolling friction (reaction) force, ${\bf f}_r$, 
applied to the particle at a point slightly in front of the normal to the surface from 
the particle's center~\cite{brilliantov2,domenech,ehrlich,ristow2}. 
We note that this reaction force is actually our usual normal
force, ${\bf F}_N$. While we include the rolling friction in the 
discussion, we neglect the small modification of the normal force due to the effect of 
rolling friction.

\subsubsection{Rolling without sliding and without rolling friction}
\label{sec:no_slip}

In this work, we ignore the complex nature 
(see e.g.~\cite{rabinowicz,baumberger}) of the friction
force, and assume that there is a 
single contact point between a particle and the substrate, with the
friction force, ${\bf f}$, acting on the particle in the plane
of the substrate, in the direction given by Newton's law.  
In order to calculate the acceleration of the particle, we use the simple method
given in~\cite{gersten}.   The approach is outlined here, since in
the later sections we will use the same idea in the more complicated
settings.

If the substrate itself is moving, the friction force 
\be
{\bf f} = m {\bf a}
\label{eq:fbs}
\ee
is responsible for the momentum transfer from the substrate
to the particle, where ${\bf a}$ is
the particle acceleration.  This force produces a torque (see Fig. 3) 
\be
{\bf T} = -R\, {\bf \hat k}\times {\bf f} = I{\bf \dot \Omega}\ ,
\label{eq:torque}
\ee
where ${\bf \dot \Omega}$ is the angular acceleration of the particle.  
Assuming that there is no sliding, the velocity of the contact point
is  equal to the velocity of the substrate, ${\bf v}_S$ (this
constraint will be relaxed in section~\ref{sec:slip}, in order
to model the more general case of rolling and/or sliding)
\be
{\bf v}_S = {\bf v} + R\, {\bf \hat k}\times {\bf \Omega}\ .
\label{eq:vel}
\ee
Multiplying Eq.~(\ref{eq:torque}) by ${\bf \hat k}\times$ and using Newton
law, we obtain 
\be
{\bf a} = {I\over mR}\, {\bf \hat k}\times {\bf \dot \Omega}  \ .
\label{eq:acc}
\ee
Taking a time derivative of Eq.~(\ref{eq:vel}), and combining 
with Eq.~(\ref{eq:acc}), one obtains the 
following result for the acceleration of the center of the mass of
the particle
\be
{\bf a} = {1\over  1 + {m R^2\over I}}{\bf a}_S\ .
\label{eq:bs1}
\ee
Since, for a solid spherical particle, $I={2/5}\, m R^2$, we obtain 
${\bf a} = {2/7}\, {\bf a}_S$.  So, the acceleration of a solid particle 
moving without rolling friction, or sliding, on a horizontal surface, 
is $2/7$ of the acceleration of the surface, ${\bf a}_S$~\cite{gersten}.
\begin{figure} 
\centerline{\psfig{figure=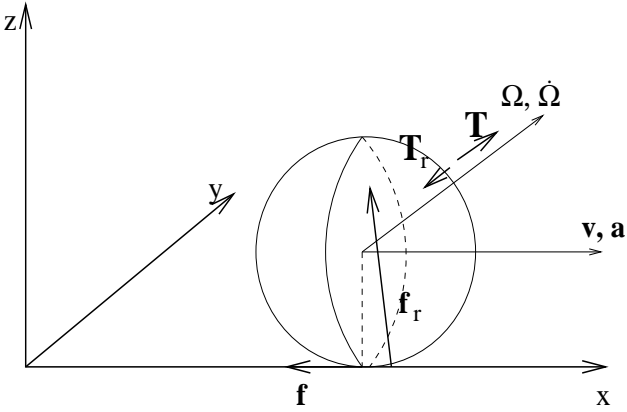,height=2.15in,width=3.25in}}
\vspace{0.1in}
\caption{The forces and torques resulting from particle-substrate interaction.
The friction force, ${\bf f}$, produces the
torque, ${\bf T}$, in the direction of the angular acceleration of the particle; 
the rolling friction force, ${\bf f}_r$, produces the 
torque, ${\bf T}_r$, in the opposite direction, so that it
leads to the decrease of the particle angular velocity, ${\bf \Omega}$.  
The deviation of ${\bf f}_r$ from the ${\bf \hat k}$ direction has been greatly 
exaggerated for the case of hard (e.g. metal) spherical particles.
}
\end{figure}

\subsubsection{Rolling without sliding with rolling friction}

The rolling friction leads to the additional force, responsible for slowing
down a particle on a surface.  As already pointed out,
this force produces the torque, ${\bf T}_r$, (see Fig. 3),
in the direction opposite to the angular velocity of the particle, ${\bf \Omega}$. 
The origins of this force are still being discussed.  The effects such
as surface defects, adhesion, electrostatic interaction, etc., that
occur at the finite contact area between the particle and the 
substrate~\cite{fuller}, as well as viscous dissipation in the bulk of 
material~\cite{tabor,green,brilliantov2} have been shown to play a role.
Fortunately, for our purposes, we do not have to understand the details 
of this force, except that it decreases the relative velocity 
${\bf{\bar v}} = {\bf v} - {\bf v}_S$ of the particle 
with respect to the substrate. The acceleration of the
particle due to this force, ${\bf a}_R$, is given by

\be
{\bf a}_R = - {\mu_r(|{\bf \bar v}|) |{\bf F}_N|\over m} {\bf{\hat{\bar v}}}\ ,
\label{eq:roll}
\ee
where the coefficient of rolling friction, $\mu_r(|{\bf \bar v}|)$, is
defined by this equation, 
${\bf{\hat{\bar v}}}={{\bf{\bar v}}/ |{\bf{\bar v}}|}$, and
${\bf F}_N$ is the normal force.  Alternatively, one could define the
coefficient of rolling friction as the lever hand of the reaction force
${\bf f}_r$ shown in Fig. 3~\cite{domenech}; for our purposes, the straightforward
definition, Eq.~(\ref{eq:roll}), is more appropriate.
For the case of steel spherical particles rolling on a copper
substrate, the typical values of $\mu_r$ are of the order of 
$10^{-3}$~\cite{golob,behringer}.  Realistic modeling of the experiments where
rolling friction properties are of major importance (such as recent
experiment~\cite{behringer}, which explores a system consisting of two kind of
particles, distinguished by their rolling friction) requires accounting for
velocity dependence of $\mu_r = \mu_r (|{\bf \bar v}|)$. 

We note that there is an additional frictional force, which slows down
the rotations of the particles around their vertical axes.  While
this additional force is to be included in general, we choose to
neglect it here, since for the experimental situation
in which we are interested~\cite{golob,behringer}, the collisions between
particles occur on a time scale that is much shorter than the
time scale on which this rotational motion is considerably slowed
down by the action of this frictional force (for other experimental 
systems, e.g. rubber spheres, this approximation would be unrealistic).

To summarize, a particle rolling without sliding on a horizontal surface
experiences two kind of forces: first, the surface transfers momentum
to the particle, ``pulling'' it in the direction of its own motion and
leading to the acceleration, ${\bf a}$, given by Eq.~(\ref{eq:bs1}). 
Second, due to the rolling friction, the particle is being
slowed down, i.e. it is being accelerated with the acceleration, ${\bf a}_R$, 
in the direction opposite to the relative velocity of the particle 
and the substrate. 

\subsubsection{Rolling with sliding}
\label{sec:slip}

Finally we are ready to address the problem of sliding.  
Sliding of a particle that is rolling on a substrate occurs when the
magnitude of the friction force, resulting from Eq.~(\ref{eq:fbs}),
reaches its maximum allowed value $|{\bf f}_{max}|$, where
\be
|{\bf f}_{max}| = \mu_s |{\bf F}_N|\ .
\label{eq:cond_slip}
\ee
Here ${\bf F}_N$ is the normal force with which the substrate acts on
a particle in the perpendicular, ${\bf \hat k}$, direction, and
$\mu_s$ is the coefficient
of static friction between a particle and the substrate.
Once the condition~(\ref{eq:cond_slip}) is satisfied, the friction
force has to be modified, since now this force arises not from the 
static friction, but from the kinematic one.  The direction of ${\bf f}$ 
is opposite to the relative (slip) velocity of the contact point of a particle and 
the substrate, ${\bf{\bar u}}$.  Here ${\bf{\bar u}} = {\bf u} - {\bf v}_S$,
where ${\bf u}$ is the velocity of the contact point.
The magnitude of ${\bf f}$ is equal to the product of the normal force and the 
coefficient of kinematic (sliding) friction, $\mu_k$
\be
{\bf f} = -\mu_k |{\bf F}_N| {\bf{\hat{\bar u}}}\ .
\label{eq:force_slip1}
\ee
The typical range of values of $\mu_s$ and $\mu_k$ 
are $0.5-0.7$ and $0.1-0.2$, respectively.
In the subsequent analysis we neglect rolling friction, 
since the rolling friction coefficient, $\mu_r$, is two orders of magnitude 
smaller than both $\mu_s$ and $\mu_k$.

{\it The condition for sliding if there are no collisions}.
In this simple case, the friction force is given by 
Eqs.~(\ref{eq:fbs},~\ref{eq:bs1}).  From the
the condition for sliding, $|{\bf f}| = |{\bf f}_{max}|$, 
where $|{\bf f}_{max}|$ is given by Eq.~(\ref{eq:cond_slip}), we obtain
that the sliding occurs if
\be
|{\bf a}_S| \ge  \left( 1+{mR^2\over I}\right) {\mu_s\over m} |{\bf F}_N|\ .
\label{eq:acc_subs}
\ee
As expected, if the substrate is accelerated with large acceleration,
a particle slides.  On a horizontal surface, the condition 
for sliding is $|{\bf a}_S| > |({\bf a}_S)_{min}|$, where
$|({\bf a}_S)_{min}| = (1+{mR^2/ I})\, \mu_s g$.  For a solid steel sphere, 
$\mu_s \approx 0.5$, so $|({\bf a}_S)_{min}|\approx 1.75\, g$.  We note
that this result does not depend on the diameter of a particle.

\section{Motion of the particles}\label{sec:move}

The preceding section gives the results for the forces that
the particles experience, because of their collisions, as well
as because of their interaction with the substrate.  Now we
consider the mutual interaction of these effects and
give expressions that govern the motion of the particles. 

Similarly to before, we consider first the case 
where the particles roll without sliding.  Sliding is included
in the second part of the section.

\subsection{Motion without sliding}

In this section, we do not include rolling friction,
since its effect is rather weak compared to the effects due to the collisions 
and the substrate motion.  It is important to note
that this approximation is valid only during the collisions;
in between of the collisions, the rolling friction force has to be included, 
since it is the only active force other than gravity.

The linear acceleration of a particle (in $x-y$ plane) is given by
\be
m {\bf a} = {\bf F}_N^c + {\bf F}_S^c + {\bf f} + m{\bf a}_G  \ ,
\label{eq:accs}
\ee
where ${\bf F}_N^c$ and ${\bf F}_S^c$ are the forces on a
particle due to the collision, in the normal and tangential directions,
respectively; ${\bf f}$ is the friction force, and ${\bf a}_G$ 
is the acceleration due to gravity.  Figure 2 shows a simple 2D example, where, 
for clarity, ${\bf F}_S^c$, the rolling friction force, and the rotations, 
characterized by $\Omega^z$, are not shown.
The torque balance (generalization of Eq.~(\ref{eq:torque}))
implies that the angular acceleration of a particle is given by 
\be
{\bf \dot \Omega} = -{R\over I} ({\bf \hat k}\times {\bf f}
+ {\bf \hat n}\times {\bf F}_R^c)\ ,
\label{eq:ang_accs1}
\ee
where we concentrate only on the rotations in the $x-y$ plane 
(the rotations characterized by $\Omega^z$ enter into the
definition of ${\bf F}_R^c$ only).  Following the same approach that led 
to Eq.~(\ref{eq:bs1}), one obtains the result for the linear acceleration
\be
m {\bf a} = {{m R^2\over I} 
[{\bf F}_N^c + {\bf F}_S^c + ({\bf \hat k}\cdot {\bf F}_R^c ){\bf \hat n}]
+ m({\bf a}_S+{\bf a}_G)\over 1+{m R^2\over I}} \ .
\label{eq:acc_no_slip}
\ee
Similarly, one can solve for the friction force, ${\bf f}$,
that the substrate exerts on a particle
\be
{\bf f} = - {{\bf F}_N^c + {\bf F}_S^c + m({\bf a}_G-{\bf a}_S) -
{m R^2\over I}({\bf \hat k}\cdot {\bf F}_R^c ){\bf \hat n}\over
 1+{m R^2\over I}} \ .
\label{eq:fric_force}
\ee
For a solid particle, one obtains
\be
m {\bf a} = {5\over 7} [{\bf F}_N^c + {\bf F}_S^c + 
({\bf \hat k}\cdot {\bf F}_R^c ){\bf \hat n}] +
{2\over 7}\, m ({\bf a}_S+{\bf a}_G) \ .
\label{eq:acc_no_slip1}
\ee

Eqs.~(\ref{eq:acc_no_slip},~\ref{eq:acc_no_slip1})
effectively combine the acceleration due to 
the substrate motion, Eq.~(\ref{eq:bs1}), and the acceleration due to the 
collisions, Eq.~(\ref{eq:coll}).  We note that, due to the interplay between
angular and linear motion of the particles, {\it the total acceleration},
given by Eq.~(\ref{eq:acc_no_slip}), {\it is not simply the sum of the
accelerations due to the collisions, gravity, and the frictional
interaction with the substrate.} 
These interactions are 
effectively coupled, and one should not consider them separately.

\subsection{Motion with sliding}\label{sec:slip_motion}

The condition for sliding follows immediately from
Eq.~(\ref{eq:fric_force}), and from the sliding condition,
$|{\bf f}| = |{\bf f}_{max}|$, yielding
\be
{\left | {\bf F}_N^c  + {\bf F}_S^c
-{m R^2\over I} ({\bf \hat k}\cdot {\bf F}_R^c ){\bf \hat n} + 
m({\bf a}_G-{\bf a}_S)\right |\over 1 + {m R^2\over I}} =
\mu_s |{\bf F}_N| \ .
\label{eq:slip_res}
\ee
If this condition is satisfied,
then ${\bf f}$ is given by Eq.~(\ref{eq:force_slip1}).

Let us concentrate for a moment on the condition for sliding of a solid particle,
on a horizontal static substrate, and neglect 
${\bf F}_R^c$ and ${\bf F}_S^c$. The sliding condition
is now given by $|{\bf F}_N^c| = (1+{mR^2/I})\, \mu_s |{\bf F}_N|$.
As one would expect, this result resembles the condition for
sliding due to the motion of the substrate, Eq.~(\ref{eq:acc_subs}), since the
two considered situations are analogous (the acceleration of the surface
${\bf a}_S$, plays the same role as the collision force, ${\bf F}_N^c$, 
scaled with the mass of a particle).  Consequently, if the substrate is being 
accelerated in the direction of ${\bf F}_N^c$, a larger ${\bf F}_N^c$ is required
to produce sliding.  {\it So, it is actually the relative acceleration of a 
particle with respect to the substrate motion which is relevant in determining 
the condition for sliding}.  

The effect of  ${\bf F}_R^c$ on the sliding condition is more involved.
Since ${\bf F}_N$ is connected with ${\bf F}_R^c$
via Eq.~(\ref{eq:fn_mod}), ${\bf F}_R^c$ modifies both sides of Eq.~(\ref{eq:slip_res}). 
The net effect of ${\bf F}_R^c$ is discussed in some
details in section~\ref{sec:discussion}.

If the sliding condition, Eq.~(\ref{eq:slip_res}), is satisfied, one has to
relax the no-slip condition,  Eq.~(\ref{eq:vel}).  Instead of no-slip
condition, we have
\be
{\bf u} = {\bf v} + R\, {\bf \hat k}\times {\bf \Omega}\ .
\label{eq:slip}
\ee
If ${\bf u} = {\bf v}_S$, the sliding velocity,
${\bf{\bar u}} = {\bf u}- {\bf v}_s$, vanishes.  
The equation for the ``sliding acceleration'', ${\bf \dot {\bar u}}$
(relative to the acceleration
of the substrate, ${\bf a}_S$), 
follows similarly to Eq.~(\ref{eq:acc_no_slip}), 
\begin{eqnarray}
m {\bf \dot {\bar u}} = && \left ( 1+{m R^2\over I}\right ) {\bf f} 
+{\bf F}_N^c + {\bf F}_S^c - {m R^2\over I}({\bf \hat k}\cdot {\bf F}_R^c ){\bf \hat n}
\nonumber \\ && +m ({\bf a}_G-{\bf a}_S)\ .
\label{eq:slip_acc}
\end{eqnarray}

To summarize, the linear acceleration of a particle 
is given by Eq.~(\ref{eq:accs}), where the normal force, ${\bf F}_N^c$, 
is given by Eq.~(\ref{eq:fn}) and the tangential force, ${\bf F}_S^c$, 
by Eq.~(\ref{eq:shear}).  If there is no sliding, then 
the friction force, ${\bf f}$, is given by 
Eq.~(\ref{eq:fric_force}); on the other hand, if the sliding condition,
Eq.~(\ref{eq:slip_res}), is satisfied, ${\bf f}$ is given by
Eq.~(\ref{eq:force_slip1}).   Further, ${\bf a}_S$ is the 
acceleration of the substrate, and ${\bf a}_G$, the acceleration due to 
gravity, is given by Eq.~(\ref{eq:grav}). The angular acceleration 
of a particle, ${\bf \dot \Omega}$, is given by Eq.~(\ref{eq:ang_accs1}), 
where the rotational force due to the collision, ${\bf F}_R^c$, is given by 
Eq.~(\ref{eq:force_r}).  Finally, the sliding acceleration, 
${\bf \dot {\bar u}}$, follows from Eq.~(\ref{eq:slip_acc}).  
As mentioned earlier, the acceleration due to rolling friction, ${\bf a}_R$, 
is not important as long as much stronger collision, or friction
forces, are present; it has to be added to the linear acceleration, ${\bf a}$, 
for realistic modeling of the motion of the particles between the collisions.
The rotational motion of the particles, characterized by
$\Omega^z$, enters into the model only by modifying the collision
force between the particles in the tangential direction, ${\bf F}_S^c$. 

The general expressions given in this section are 
used in the MD type simulations~\cite{we},
in order to simulate the motion of a set of particles on an inclined
plane.  In this paper, we apply the results to a simple setting, and
obtain the analytic results which provide better insight into  
the relative importance of various interactions.  This is the subject
of the next section.

\section{Discussion}\label{sec:discussion}

The analysis of the preceding section gives rather general results, 
that provide all the information needed for modeling of the particles' motion.
On the other hand, the complexity of the final results obscures simple
physical understanding.  In this section, we concentrate on the particular case
explored in recent experiments~\cite{golob,behringer}, performed
with steel particles on a metal substrate, and choose parameters appropriate
to this situation.  This system allows for significant simplifications, 
so that we are able to obtain rather simple analytic results.  The assumptions
which we use in what follows are summarized here for clarity:
\begin{itemize}
\item Particles move just in one, ${\bf \hat i}$ direction;
\item Particles are rolling without sliding prior to a collision;
\item The relative velocity of the particle prior to a collision,
$v^0_{rel} = |{\bf v}_i^0 - {\bf v}_j^0|$, is
assumed to be in the range $100$ cm/s$> v^0_{rel}> 0.1$ cm/s.  It is assumed that
the linear force model, Eq.~(\ref{eq:fn}), is
appropriate for these velocities, but a nonlinear model is used to
determine the approximate expressions for the force constant, $k$,
and the damping parameter, $\gamma_N$ (Appendices~\ref{sec:linear}
and~\ref{sec:nonlinear}).  For smaller relative velocities, we will see that the 
interaction with the substrate substantially complicates the analysis.  Still, this range of 
velocities is the most common one in the experiments~\cite{golob,behringer}, so we do not 
consider that this is a serious limitation;
\item The particles are assumed to be moving on a horizontal, static
substrate; the analysis could be easily extended to other situations;
\item We neglect rolling friction, since its effect is negligible during
a collision, or as long as the particles slide;
\item For simplicity of the presentation, we assume that the particles initially move 
with the velocities in the opposite directions; the final results are
independent of this assumption.  To avoid confusion, the particle $i$ is always assumed 
to be initially either static, or moving in the $-{\bf \hat i}$ direction.
\end{itemize}

We are particularly interested in answering the following questions:
\begin{itemize}
\item What is the condition for sliding to occur;
\item How long does a particle slide after a collision, and what is the distance
traveled by a particle during this time;
\item How much of the translational energy and linear momentum of a particle 
are lost due to sliding.
\end{itemize}

In order to fully understand the problem, we start with the simplest possible
situation, and that is a symmetric collision of two particles moving with 
same speeds, but opposite velocities.  Further, we assume the collision
to be totally elastic. From this simple example, we conclude that the
particle-substrate interaction is not important {\it during} a collision, at least not
for the before mentioned range of particle velocities.
Next we look into the case of more realistic, inelastic collision.  
Finally, we consider general inelastic, asymmetric collision of two particles. 

\subsection{Symmetric collisions}\label{sec:symmetric}

Let us concentrate on the first part of a symmetric central  
collision of two particles, and analyze the forces acting on 
the particle $i$, as shown in Fig. 4a.  The only collision force 
acting on the particle is ${\bf F}_N^c$, the substrate is static and horizontal, 
and the rolling friction can be neglected.  In this case, 
Eqs.~(\ref{eq:ang_accs1},~\ref{eq:acc_no_slip},~\ref{eq:slip_acc}) simplify to
\begin{eqnarray}
m {\bf a} =&& {\bf F}_N^c + {\bf f}
\label{eq:accs1_s}\ , \\
{\bf \dot \Omega} =&& -{R\over I} {\bf \hat k}\times {\bf f}
\ ,
\label{eq:ang_accs1_s} \\
m {\bf \dot u} =&& \left ( 1+{m R^2\over I}\right ) {\bf f}
 + {\bf F}_N^c \ .
\label{eq:slip_acc_s}
\end{eqnarray}
Further, the friction force is given by
\begin{eqnarray}
\displaystyle{{\bf f}}=
\left\{ \begin{array}{ll}
\displaystyle{- {{\bf F}_N^c \over \left ( 1 + {mR^2\over I}\right)}
} & \mbox {if}  \ 
|{\bf f}|< \mu_s |{\bf F}_N|  \\
\displaystyle{ -\mu_k|{\bf F}_N| {\bf {\hat u}} 
} & \mbox {otherwise}  \ 
\end{array}
\right. 
\label{eq:e.00}
\end{eqnarray}
Let us assume that the sliding condition, ${\bf f} = \mu_s |{\bf F}_N|$, is
satisfied.  By inspecting Eq.~(\ref{eq:slip_acc_s}), we observe that 
the first term on the right hand side is the one that decreases 
the sliding acceleration continuously, and possibly brings the particle back to 
pure rolling, since it acts always in the direction opposite to the sliding velocity.  
Because of the constraint on the friction force given by Eq.~(\ref{eq:e.00}), 
and $\mu_k < \mu_s$, the right hand side of Eq.~(\ref{eq:slip_acc_s}) gives 
a net contribution in $+{\bf \hat i}$ direction.  So, when sliding begins, 
the particle $i$ experiences the sliding acceleration in the ${\bf \hat i}$ 
direction, leading to the sliding velocity, ${\bf u}$, in the same direction, 
as shown in Fig. 4a.  In other words, for this situation, the particle $i$
is still moving to the left, with angular velocity in the $-{\bf \hat j}$ direction,
but it is sliding to the right, with sliding velocity ${\bf u} $. 
Let us also note that ${\bf f}$ slows down the angular motion of the particle, 
as it can be seen by inspection of Eq.~(\ref{eq:ang_accs1_s}).  
Next, we consider the typical situation during the second part of the collision, 
when the particles are moving away from each other (Fig. 4b).  
Analysis shows that almost all of the conclusions about the situation 
depicted in Fig.~4a extend to this situation; in particular the directions 
of the sliding velocity, ${\bf u}$, and the friction force, ${\bf f}$, are the same.  
\begin{figure}
\centerline{\psfig{figure=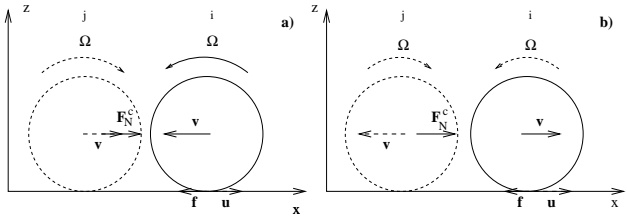,height=1.1in,width=3.25in}}
\vspace{0.1in}
\caption{The forces considered in this section
($x-z$ plane).  Part a) shows
the first part of the collision, when the particles are still moving
towards each other, and b) shows the  second part of the 
collision.   For clarity, only forces on the particle $i$ are shown.
The rotational motion is discussed in the text.
}
\end{figure}

After understanding this basic situation, it is easier to understand
the role of the remaining terms, given in section~\ref{sec:slip_motion}, but ignored in
Eqs.~(\ref{eq:accs1_s}-\ref{eq:e.00}).  The contributions from
gravity and the motion of the substrate are obvious.
The analysis of the collision force in the tangential direction, ${\bf F}_S^c$, 
is similar to the one about ${\bf F}_N^c$, since the forces in the normal 
and tangential directions could be considered independently.
The contribution coming from ${\bf F}_R^c$ is discussed in the following
sections.  We note that in the case shown in Fig. 4 (the particles
are initially moving with exactly opposite velocities), the contribution
from ${\bf F}_R^c$ vanishes, since it is proportional to the relative
velocity of the point of contact in the ${\bf \hat k}$ direction.

\subsubsection{Symmetric elastic collision}
\label{sec:symmetric_elastic}

Let us define the compression by $x={(d - r_{i,j})/d}$, where
$r_{i,j}$ is the distance between the centers of colliding particles, and $d$
is the particle diameter ($x>0$ is required if a collision occurs).
The maximum compression is given by (Appendix~\ref{sec:linear})
\be
x_{max}^0= { v^0_{rel}\over d \omega_0}\ ,
\label{eq:xmax_0}
\ee 
where, for a symmetric collision, the relative velocity $v^0_{rel} = 2 v^0$, 
and $v^0$ is the initial speed of the particles.   For simplicity,
we use scalar notation and the sign convention that +sign of the translational
velocities refers to the motion in $+{\bf \hat i}$ direction, and +sign
of the angular velocities/accelerations to the rotations in $+{\bf \hat j}$ 
direction.  In obtaining Eq.~(\ref{eq:xmax_0}), it was assumed that only the collision 
forces are important in determining $x_{max}^0$.  More careful analysis 
given in Appendix~\ref{sec:substrate} provides justification for this assumption. 
The natural frequency, $\omega_0$,  associated with the linear force model specified by 
Eq.~(\ref{eq:fn}), is given by $\omega_0^2 = {2 k/m}$.  It is related to the duration of the 
collision via $t_{col} = \pi/\omega_0$.  The nonlinear force model 
(see Appendix~\ref{sec:nonlinear}), predicts that $\omega_0$ very weakly depends on $v^0_{rel}$;
we observe that typically $\omega_0\approx 10^{-5}$ s, so that for $v^0_{rel}\approx 10$ cm/s, 
$x_{max}^0\approx 2\times 10^{-4}$. Figure 5a shows the dependence of
$x_{max}^0$ on the relative initial velocity of the particles.  

It is of interest to estimate the range of the compression depths for which
the sliding condition, $|{\bf f}| = \mu_s|{\bf F}_N|$, 
is satisfied, leading to sliding of the particles with respect to the
substrate. Sliding occurs when $|{\bf F}_N^c| \ge ({1+mR^2/I})\, \mu_s m g$
(see Eqs.~(\ref{eq:accs1_s}-\ref{eq:e.00})).  Using the expression for the normal force, 
Eq.~(\ref{eq:fn}), we obtain that this condition is satisfied for 
$x_{max}^0\ge x_{slip} \ge x_{min}^0$, where (see Appendix~\ref{sec:approx})
\be
x_{min}^0 = \left ( 1+{mR^2\over I}\right ) 
{\mu_s g\over d \omega_0^2} \ .
\label{eq:xmin_0}
\ee
Using the values of the parameters as specified in Appendices~\ref{sec:nonlinear} 
and \ref{sec:approx}, we note that for $v^0_{rel}\approx 10$ cm/s, $x_{min}^0
\approx 3\times 10^{-7}$.  For $d = 4$ mm, $x_{min}^0$ is on atomic length scale,
so we conclude that a particle slides during almost all of the course of a symmetric,
elastic collision.  The results for $x_{min}^0$ are shown in Fig. 5b.
The dependence of $\omega_0$ on $v^0$ (see Appendix~\ref{sec:nonlinear}),
leads to the increased values of $x_{min}^0$ as $v^0\rightarrow 0$.

The compression, $x_{min}^0$, is reached at time $t_{min}^0$, measured from the 
beginning of the collision (see Appendix~\ref{sec:approx})
\be
t_{min}^0= \left (1 + {m R^2\over I}\right)  
{ \mu_s g\over v^0_{rel} \omega_0^2}\ .
\label{eq:t_min0}
\ee
For the choice of parameters as given in Appendix~\ref{sec:nonlinear}, we
obtain $t_{min}^0\approx 10^{-8}$, and ${t_{min}^0/t_{col}}\approx 
5\times 10^{-4}$, confirming our conclusion that sliding with the respect to the substrate 
is the dominant motion of the particles during a symmetric, elastic collision. 

From Eq.~(\ref{eq:xmin_0}) we can also deduce under what conditions sliding
occurs.  Obviously, we require that $x_{min}^0< x_{max}^0$.  Using 
Eqs.~(\ref{eq:xmax_0},~\ref{eq:xmin_0}), we obtain that the initial 
velocities of the particles have to satisfy $v^0_{rel}\gg v^b$, where
(see Appendix~\ref{sec:approx})
\be
v^b = \left( 1 + {mR^2\over I}\right)  {\mu_s g \over \omega_0} \ .
\label{eq:vel_b_elastic}
\ee
For the set of parameters given in Appendix~\ref{sec:nonlinear}, this
expression yields very small value, $v^b\approx 10^{-2}$ cm/s.  
So, sliding occurs during almost all collision occurring in typical
experiments~\cite{golob,behringer}.

Let us now look into the rotational motion of the particles during a 
collision.  The friction force is the only one which produces angular acceleration.
Without loss of generality, we consider the particle $i$, which is
assumed to move initially in $-{\bf \hat i}$ direction, with angular
velocity $\Omega_i^0=-{v^0/R}$.  Integrating over the duration of the collision
gives the result for the angular velocity of the particle at the end of the 
collision (at $t=t_{col}$) (see Appendix~\ref{sec:rotations})
\begin{eqnarray}
\Omega_i^{f0} &&= \Omega_i^0 +{m R\over I}{g\over \omega_0}\times \nonumber  \\
&& 
 \left[ \pi \mu_k + \left( 1 + {mR^2\over I}\right) 
{\mu_s g\over v^0_{rel} \omega_0} 
\left( {1\over 2}\mu_s - \mu_k\right) \right]\ .
\label{eq:omega_0}
\end{eqnarray}
The sliding velocity of the particle $i$ at $t=t_{col}$, 
$u_i^{f0}=v^0-R\Omega_i^{f0}$ (see Eq.~(\ref{eq:slip})), now follows
\begin{eqnarray}
u_i^{f0} &&= v^0_{rel} - {m R^2\over I}{g\over \omega_0}\times \nonumber \\
&&  \left[ \pi \mu_k + \left( 1 + {mR^2\over I}\right)
{\mu_s g\over 2 v^0_{rel} \omega_0} \left( {1\over 2}\mu_s - 
\mu_k\right) \right]\ .
\label{eq:uslip_0}
\end{eqnarray}
This is important result, since the sliding velocity of a particle 
at the end of a collision determines the energy and momentum loss 
due to the sliding after the collision.  Using the parameters as in Appendix~\ref{sec:nonlinear}, 
we note that the contribution of the second term in Eq.~(\ref{eq:uslip_0})
is approximately $ 0.01$ cm/s. So, we conclude that the frictional interaction of a particle 
with the substrate {\it during} a collision only very weakly influences the sliding velocity of
a particle at the end of the collision.  Similarly, the angular velocity
is only slightly modified, as shown in Fig. 4.  {\it A particle exits an elastic, symmetric
collision with the angular velocity which is almost equal to its initial angular
velocity, resulting in the sliding velocity equal to twice of its initial
translational velocity.}

The fact that the particle-substrate interaction is negligible during a collision
follows also from a simple energy argument.  Figure 5 
shows that the maximum compression depth is of the order of $10^{-4}$, in
units of the particles diameter.  So, the order of magnitude of the ratio of the 
energies involved in the particle-substrate interaction, $E_{p-s} \approx \mu_s m g d x_{max} $, 
and of the energy involved in the collision itself, $E_{coll}\approx {k(d x_{max})^2/2}$, is given by 
\be
{E_{p-s}\over E_{col}} \approx {\mu_s m g v^0_{rel} t_{col}\over m (v^0_{rel})^2} \approx
{g t_{col}\over v^0_{rel}}\nonumber \ ,
\ee
where Eq.~(\ref{eq:xmax_0}) has been used.  For $v^0_{rel}= 10$ cm/s, we obtain
${E_{p-s}/ E_{col}}\approx 2\times 10^{-3}$.  Clearly, for all collisions 
characterized by very short collision times (equivalently, small maximum 
compression depths), this ratio is very small number, 
assuming common particle velocities.  Correspondingly, the particle-substrate interaction
{\it during} a collision influences very weakly the dynamics.
Considerable modification of this estimate could be expected in the case of
``softer'' collisions, where both the duration of a collision and maximum
compression depth are much larger.

\subsubsection{Symmetric inelastic collision}
\label{sec:symmetric_inelastic}

Inelasticity of a collision introduces damping parameter,
$\gamma_N$, which is related to the material 
constants in Appendices~\ref{sec:linear} and~\ref{sec:nonlinear}.  
The damping is directly connected with the coefficient of restitution $e_n$
by $\gamma_N = -{2/t_{col}} \ln (e_n)$ (Appendix~{\ref{sec:nonlinear}).  
The collisions of steel spheres are rather elastic 
(typically $e_n\approx 0.9$), so we are able to introduce
a small parameter, $\epsilon = {\gamma_N/ \omega_0}\approx - {2/\pi}(1-e_n)\ll 1$.
In what follows we perform consistent perturbation expansions of the equations
of motion, and include only the corrections of the order 
$O(\epsilon)$.  For the completeness, we
also include the terms due to the interaction with the substrate, even though we have 
already shown that this interaction is not of importance for the physical 
situation we are interested in.

The maximum compression is now given by (see Appendix~\ref{sec:linear})
\be
x_{max}= {v^0_{rel}\over d \omega_0} \left( 1-{\pi\over 2}\epsilon
 +O(\epsilon^2) \right)\ .
\label{eq:max_comp}
\ee
Figure 5a shows the result for $x_{max}$, for
a few values of $e_n$, using the parameters given in 
Appendix~\ref{sec:nonlinear}. 
In Appendix~\ref{sec:nonlinear} it is shown that the $\omega_0$ and $t_{col}$
are weakly dependent on the initial velocity ($t_{col}\sim (v^0)^{-1/5}$), 
so that the results for the maximum compression scale with the initial velocity
as $x_{max}\sim (v^0_{rel})^{4/5}$, resulting in the slight curvature of 
the $x_{max}$ curves in Fig. 5a.  In Appendix~\ref{sec:approx} it is shown that for 
inelastic collisions, the time, $t_{min}$, at which sliding starts goes to zero, since the
corrections due to damping are typically stronger than corrections due to 
the particle-substrate interaction.  Consequently, the angular acceleration,
given by Eq.~(\ref{eq:ang_accs1_s}),
is constant during whole course of the collision.  For the particle $i$,
$ \dot \Omega_i=\mu_k g\, {mR/I} $, so that at the end of the
collision ($t=t_{col}$)
\be
\Omega^f_i = \Omega^0_i +{m R\over I} {\mu_k g\pi\over \omega_0} \approx \Omega^0_i \ ,
\ee
where $\Omega^0_i = - {v^0/R}$.  The sliding velocity of the particle $i$ 
at $t=t_{col}$ is given by
\be
u^f_i = {1\over 2}(1 + e_n) v^0_{rel} - 
{m R^2\over I}{\mu_k g\pi\over \omega_0}\approx 
{1\over 2}(1 + e_n) v^0_{rel}\ .
\label{eq:u_f}
\ee
Similarly to the discussion following Eq.~(\ref{eq:uslip_0}), we observe that 
the friction during a collision leads to negligible corrections.  In what follows, 
we ignore these corrections, and assume
$\Omega^f_i = \Omega^0_i$, and $u^f_i = {(1 + e_n) v^0_{rel}/2}$.
\begin{figure}[htb]
\centerline{\psfig{figure=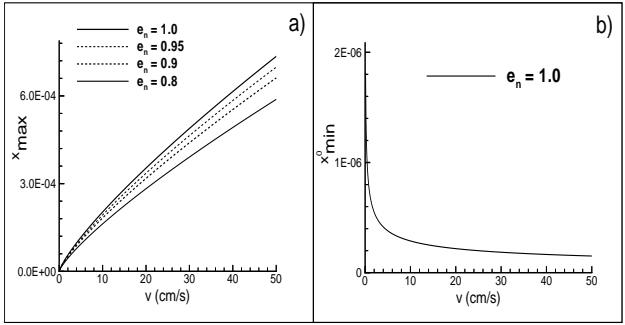,height=1.7in,width=3.25in}}
\vspace{0.1in}
\caption{a) The maximum compression, $x_{max}$; b) the minimum
compression, $x_{min}^0$, required to produce sliding.  Both
quantities are scaled with a particle diameter.  For inelastic
collisions, $x_{min} = 0$ (see text).
}
\end{figure}

\subsubsection {Sliding after a symmetric collision}

The particle $i$ exits a symmetric collision with the translational velocity
$v^f_i = e_n v^0$ (in $+{\bf \hat i}$ direction), 
and with the sliding velocity $u^f_i$, given by Eq.~(\ref{eq:u_f}).  
After the collision, it experiences the friction force, resulting in
the sliding acceleration $\dot u_i = - ( 1+{mR^2/ I} ) \mu_k g$, 
as it follows from Eq.~(\ref{eq:slip_acc_s}), where ${\bf F}_N^c$ is now absent. 
This friction force is present as long as the sliding velocity
is nonzero.  It slows down the particle, and leads to the corresponding loss of the
translational kinetic energy and linear momentum.  Neglecting rolling friction, we obtain 
the result for the time, $t^s$, measured from the end of the collision, when sliding stops
(due to the symmetry this result is the same for both particles)
\be
t^s = {u^f_i\over |\dot u_i|} = 
{ {1\over 2}(1 + e_n) v^0_{rel} \over \mu_k g 
\left( 1+ {mR^2\over I}\right) }\nonumber \ .
\ee
The translational velocity of the particle $i$ at the time $t^s$ is given by 
\be
v^s_i = v_i (t=t^s) =  {1\over 2} v^0_{rel} \, { e_n {m R^2\over I} - 
1\over 1 + {m R^2\over I}}\ .
\label{eq:v_s}
\ee
The angular velocity of the particle $i$ at this time is
$\Omega^s_i = \Omega_i (t=t^s) = {v^s_i/R}$, since the particle does not
slide anymore.  We observe that the translational motion of the particles is considerably 
slowed down due to the friction force; for solid spheres, and $e_n=0.9$, 
$|v^s_k| \approx 0.36\, v^0$, ($k=i,j$).  Equation~(\ref{eq:v_s}) gives that
for ${I/ m R^2} > e_n$, $v^s_i$ is negative, meaning that the particle is moving
{\it backwards} at the time when sliding ceases.  For almost elastic collisions of solid
 particles this condition is not satisfied, so the particles are still moving away from the 
collision point at time $t^s$.

Until the time $t^s$, each of the particles travels the distance $s$ away from the
point where the collision has taken place, given by
\be
s = {(v^0_{rel})^2\over 8 \mu_k g \left( 1+ {mR^2\over I}\right)^2 }
(1+e_n) \left( e_n -1 + 2 {mR^2\over I} e_n\right)\nonumber \ .
\ee
Figure 6 shows the results for $t^s$ and $s$.
For $v^0_{rel} = 10$ cm/s, and $e_n = 0.9$, the particles slide
during the time $t^s \approx 0.03$ s, and  
$s\approx 0.1$ cm.  These results compare well with preliminary experiments.
More precise analysis and the comparison with experimental results
will be given elsewhere~\cite{ben}.
\begin{figure}[htb] 
\centerline{\psfig{figure=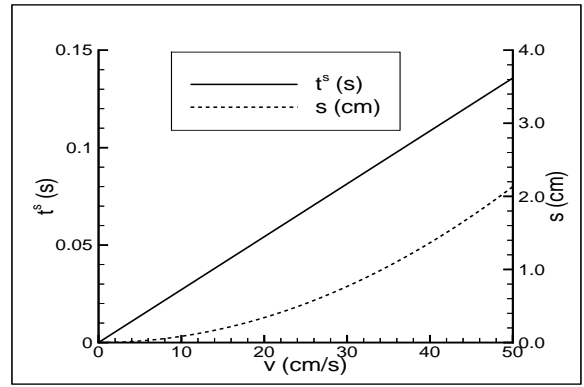,height=2.0in,width=3.0in}}
\vspace{0.1in}
\caption{The time, $t^s$, until which a particle slides after the 
collision, and the distance, $s$, traveled during this time.  Here
$v=v_0^{rel}$ is the initial relative velocity of the particles.
The parameters are as specified in the text.
}
\end{figure}

\subsubsection {The loss of the translational energy 
and momentum due to sliding in a symmetric collision}

Let us define the energy loss due to sliding, $\Delta \bar E_{slip}$, as
the difference between the translational kinetic energy of a particle 
just after it has undergone a collision, and its translational kinetic 
energy at time $t^s$ after the collision (scaled with the reduced mass) . Therefore,
\be
\Delta \bar E_{slip} = (v^f)^2 - (v^s)^2 \ .
\label{eq:eloss_s}
\ee
The relative loss of energy is defined as 
$\Delta E_{slip}={\Delta \bar E_{slip}/E_0}$,
where $E_0= (v^0)^2$.
We note that very little energy is lost due to sliding while the 
collision is taking place.  The sliding loss of energy occurs after
the collision, and it is equal to the work done by the 
friction force.  Using Eq.~(\ref{eq:v_s}), we obtain the result for 
the relative energy loss due to sliding,
\be
\Delta E_{slip} = (1+e_n){ e_n -1 + 2 e_n {mR^2\over I} \over
\left( 1+ {mR^2\over I}\right)^2} \ .
\label{eq:energy}
\ee
Figure 7a shows $\Delta E_{slip}$ for a range of values of $e_n$ (assuming
solid spheres).  In the limit of an elastic collision, $e_n \rightarrow 1$, and we obtain
$\Delta E_{slip} \approx 0.8$.  {\it So, a solid particle loses approximately 
$80\, \%$ of its initial translational kinetic energy due to sliding, in a 
completely elastic symmetric collision.}  

It is also of interest to compare the relative energy loss due to sliding,
$\Delta E_{slip}$, with the energy loss due to inelasticity of a
collision, $\Delta E_{col}$.  The latter is simply given by
$\Delta E_{col} = (1 - e_n^2)$ (we neglect the small loss of
energy due to interaction with substrate {\it during} a collision), thus
\be
{\Delta E_{col}\over \Delta E_{slip}} = 
\left( 1+ {mR^2\over I}\right)^2 {1-e_n\over  e_n -1 + 2 e_n {mR^2\over I}}\ .
\label{eq:ener_s}
\ee
The result for ${\Delta E_{col}/ \Delta E_{slip}}$ is
shown in Fig. 7b. We observe that {\it in the limit of low damping,
the sliding is the main source of energy loss.}  This conclusion
is independent of the initial particle velocity or the particle diameter.
\begin{figure}[htb] 
\centerline{\psfig{figure=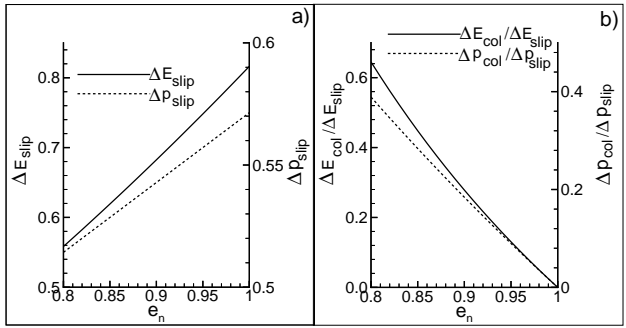,height=1.7in,width=3.25in}}
\vspace{0.1in}
\caption{a) The loss of the energy and momentum due to
sliding. b) The ratio of the loss of the mechanical energy and 
linear momentum due to
inelasticity of a collision ($\Delta E_{col},\ \Delta p_{col})$, 
and due to sliding ($\Delta E_{slip},\ \Delta p_{slip})$.
}
\end{figure}

Similarly, the linear momentum lost due to sliding
in a symmetric collision (relative to the initial momentum) is given by
\be
\Delta p_{slip} = {1+e_n\over 1 + {m R^2\over I}}\ , 
\label{eq:pslip}
\ee
so (in an elastic symmetric collision),
{\it a solid particle loses approximately $60\, \%$ of its linear
momentum because of sliding} (Figure 7a).  The ratio of the loss of the linear 
momentum due to inelasticity of the collision, defined by 
$\Delta p_{col} = (1 - e_n) $, and $\Delta p_{slip} $, is given by
\be
{\Delta p_{col}\over \Delta p_{slip}} =
\left( 1 + {mR^2\over I}\right)  {1 - e_n\over 1+e_n}\nonumber
\ .
\ee
[We note that $\Delta p_{col}$ is the loss of linear momentum 
of one particle in the lab frame; inelasticity of the collisions conserves the 
linear momentum of a pair of colliding particles in the center of mass frame.]
This result is shown in Fig. 7b. Similarly to the energy considerations, we observe 
that for small damping, {\it sliding is the main source of momentum
loss}. 

\subsection{Asymmetric collisions}\label{sec:assymetric}

Next we consider a central collision between particles moving with different speeds, 
such as the one shown in Fig. 2.  On a horizontal static substrate, the 
Eqs.~(\ref{eq:ang_accs1},~\ref{eq:acc_no_slip},~\ref{eq:slip_acc}) now simplify to
(index $i$ emphasizes that the particle $i$ is being considered)  
\begin{eqnarray}
m {\bf a}_i =&& {\bf F}_{N,i}^c + {\bf f}_i
\label{eq:accs1_sa}\ , \\
{\bf \dot \Omega}_i =&& -{R\over I} ({\bf \hat k}\times {\bf f}_i+{\bf \hat n}\times {\bf F}_{R,i}^c)
\ ,
\label{eq:ang_accs1_a} \\
m {\bf \dot u}_i =&& \left ( 1+{m R^2\over I}\right ) {\bf f}_i
 + {\bf F}_{N,i}^c - {m R^2\over I} ({\bf \hat k}\cdot {\bf F}_{R,i}^c){\bf \hat n}\ .
\label{eq:slip_acc_sa}
\end{eqnarray}
Further, the friction force is given by
\begin{eqnarray}
\displaystyle{{\bf f}_i}=
\left\{ \begin{array}{ll}
\displaystyle{- {{\bf F}_{N,i}^c - {m R^2\over I} ({\bf \hat k}\cdot {\bf F}_{R,i}^c){\bf \hat n} 
\over 1 + {mR^2\over I}}
} & \mbox {if}  \,
|{\bf f}_i|< \mu_s |{\bf F}_{N,i}|  \\
\displaystyle{ -\mu_k|{\bf F}_{N,i}| {\bf {\hat u}}_i
} & \mbox {otherwise}  \
\end{array}
\right.
\label{eq:e.00a}
\end{eqnarray}
The analysis of a symmetric collision, given in section~\ref{sec:symmetric}, shows
that the frictional interaction of the particles with the substrate during a collision
can be neglected.  We use this result in the following discussion and neglect ${\bf f}_i$ in
the analysis of the collision dynamics of an asymmetric collision.  This frictional
interaction is, of course, included in the analysis of the particles' motion after
a collision, since it is the only force acting on a particle on a static, horizontal
substrate.

Normal force is being modified due to ${\bf F}_{R,i}^c$, so that
\be
{\bf F}_{N,i} = (mg  - {\bf \hat k}\cdot {\bf F}_{R,i}^c){\bf \hat k}\ .
\label{eq:fn_a}
\ee
In Appendix~\ref{sec:rotations} it is shown that, for typical experimental velocities,
the corrections of ${\bf F}_{R,i}^c$ due to its cutoff value 
(see Eq.~(\ref{eq:force_r})), could be ignored, since the cutoff leads to $O(\epsilon^2)$ 
corrections of the final angular velocity of the particle.  So, we take 
${\bf F}_{R,i}^c$ to be given by (see Eqs.~(\ref{eq:force_r},~\ref{eq:vrelz}))
\be
{\bf F}_{R,i}^c =-{\gamma_S\over 2} m R [({\bf \Omega}_i+{\bf \Omega}_j)\cdot {\bf \hat j}]{\bf \hat k}\ ,
\label{eq:force_r1}
\ee
during the whole course of a collision.  The damping parameter, 
$\gamma_S$, is kept as a free parameter for generality
(usually it is given a value $\gamma_S = {\gamma_N/2}$~\cite{thompson}).
Only constraint on $\gamma_S$ is that ${\gamma_S/\omega_0}\ll 1$, so that the coefficient of 
restitution is close to $1$.

The force ${\bf F}_{R,i}^c$ modifies the rotational motion of the particle $i$.
In Appendix~\ref{sec:rotations} it is shown that the angular velocity of the particle
at the end of a collision ($t=t_{col}={\pi/\omega_0}$) is given by 
\be
{\bf \Omega}_i^f={\bf \Omega}_i^0 - C({\bf \Omega}_i^0 + {\bf \Omega}_j^0)\ ,
\label{eq:omega}
\ee
where $C = {\pi mR^2 \gamma_S/(2 I \omega_0)} = O(\epsilon)\ll 1$. 
Equation~(\ref{eq:omega}) is correct to the first order in $\epsilon$.
Using this result, and the translational velocity of the particle
$i$ at $t=t_{col}$ (Eq.~(\ref{eq:vfin})), we obtain the sliding velocity of the particle 
$i$ at the end of the collision
\be
{\bf u}_i^f = -{1\over 2} (1 + e_n) ({\bf v}_i^0 - {\bf v}_j^0) +
C ({\bf v}_i^0 + {\bf v}_j^0)\ .
\label{eq:uf_a}
\ee
This result generalizes Eq.~(\ref{eq:u_f}), that gives
the sliding velocity of the particles undergoing a symmetric collision (the 
particle-substrate interaction during the collision has been neglected).  The
tangential force, ${\bf F}_R^c$, leads to the last term in  Eq.~(\ref{eq:uf_a}),
modifying the sliding velocity in an asymmetric collision.  This
modification depends on $|{\bf v}_i^0 + {\bf v}_j^0|$, which measures the
degree of asymmetry in a collision.

In order to exemplify the physical meaning of these results, let us consider
for a moment completely asymmetric case: a particle moving with initial velocity
${\bf v}_j^0$ and undergoing elastic collision ($\gamma_N = \gamma_S = 0$) 
with the stationary particle $i$.  In this case, we obtain ${\bf v}_j^f = 0$,
${\bf u}_j^f = -{\bf v}_j^0$.  So, the particle $j$ is stationary immediately
after the collision, but its rotation rate is unchanged (since in the limit
$\gamma_S=0$, ${\bf F}_R^c$ vanishes, and the interaction with the substrate
has been neglected), so that it has the sliding velocity equal to the negative
of its initial velocity.  Let us now consider the particle $i$.  Its translational 
velocity and sliding velocities are the same, 
${\bf v}_i^f = {\bf u}_i^f =  {\bf v}_j^0 $, since immediately after the
collision this particle has the translational velocity equal to the
initial velocity of the particle $j$, but zero rotation rate.  

{\it ``Jumping'' of the colliding particles}.
Let us finally address the assumption that the particles are bound to move on
the substrate.  From Eq.~(\ref{eq:fn_a}) we observe that, for large
positive ${\bf \hat k}\cdot {\bf F}_R^c$, this assumption could be 
violated.  The estimate is given in Appendix~\ref{sec:jump}, where it is
indeed shown that a particle colliding with a slower particle typically 
detaches from the substrate.  Fortunately, the motion of a detached
particle in the ${\bf \hat k}$ direction is limited by very small jump
heights, so that the modifications of the results for the dynamics of 
the particles in $x-y$ plane are negligible.  On the other hand, the
fact that a particle is not in the physical contact with the substrate
during a collision simplifies the analysis of the collision dynamics,
since particle-substrate interaction is not present.  We note that 
we are not aware that detachment has been observed in the experiments
performed with steel spheres moving with moderate 
speeds~\cite{golob,behringer}.  Since this effect provides direct 
insight into a collision model, it would be of considerable interest to explore 
these predictions experimentally.

\subsubsection {Sliding after an asymmetric collision}

After a collision, the particles experience the friction force, which produces
the sliding acceleration and modifies the translational velocity. 
Figures 8-11 show the results for the time that the particles spend sliding, for the
sliding distance, and for the changes in their translational kinetic energy and linear
momentum.  All of these results depend only on the sum and difference
of the initial velocities of the particles.  We define
\begin{eqnarray}
&& v_m = ({\bf v}_i^0 - {\bf v}_j^0)\cdot {\bf \hat i}\ , \nonumber \\
&& v_p= ({\bf v}_i^0 + {\bf v}_j^0)\cdot {\bf \hat i}\ ,
\end{eqnarray}
and show the dependence of our
results on these two quantities.  Since some of the approximations involving the
rotational motion of the particles during collisions (see Appendix~\ref{sec:rotations}) 
are not valid in the limit $|v_m| \ll |v_p|$, we do not consider the case $|v_m| \approx 0$ 
(which occurs when the initial velocities of the particles are almost the same).
This is the only imposed restriction.

Using 
Eqs.~(\ref{eq:slip_acc_sa},~\ref{eq:uf_a}), we
obtain the time when sliding of the particle $i$ stops
(measured from the end of a collision)
\be
t_i^s = {\left|{1\over 2} (1 + e_n) ({\bf v}_i^0 - {\bf v}_j^0) -
C ({\bf v}_i^0 + {\bf v}_j^0)\right|\over \left ( 1+{m R^2\over I}\right )
\mu_k g}\ .
\label{eq:ts_a}
\ee
Figure 8 shows the result for the sliding time for fixed
$e_n$ and $C$, as a function of $v_m$ and $v_p$.
For $v_p = 0$, we retrieve the results for the symmetric collision, shown in Fig. 6.  
We observe that $t^s_i$ just very weakly depends on $v_p$; this dependence disappears
in the limit of zero tangential damping ($C=0$), as can be seen directly from
Eq.~(\ref{eq:ts_a}).

The translational particle velocity at $t=t_i^s$ is ${\bf v}_i^s={\bf v}_i (t = t_i^s) =
{\bf v}_i^f + {\bf a}_i t_i^s$, where ${\bf a}_i = -\mu_k g {\bf \hat u}_i^f$. 
Using Eqs.~(\ref{eq:slip_acc_sa},~\ref{eq:uf_a},~\ref{eq:vfin}), we obtain
\begin{eqnarray}
{\bf v}_i^s = {1\over 2 \left ( 1+{m R^2\over I}\right ) } 
\left[ ({\bf v}_i^0 + {\bf v}_j^0)\left( 1+{m R^2\over I} -2 C\right )\right. -\nonumber\\
\left. ({\bf v}_i^0 - {\bf v}_j^0) \left( e_n {m R^2\over I} -1\right)\right]\ .
\label{eq:slip_vel_sa}
\end{eqnarray}
During the time $t_i^s$, the particle $i$ translates for the distance $|{\bf s}_i|$ from the
collision point, where
${\bf s}_i = ({\bf v}_i^f + {\bf v}_i^s) {t_i^s/2} $. 
Figure 9 shows $|{\bf s}_i|$; contrary to the sliding time $t^s_i$, the sliding 
distance does depend on the asymmetry of a collision.  This dependence is present 
since $|{\bf s}_i|$ is a function of both translational and sliding velocities of
the particle $i$.  On the other hand, $t^s_i$ depends only on
the sliding velocity of the particle.
\begin{figure}[htb]
\centerline{\psfig{figure=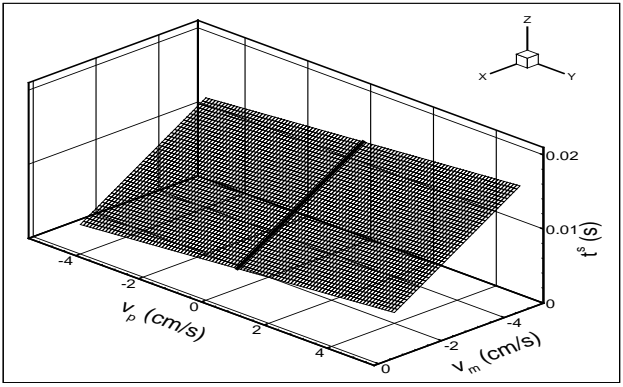,height=2.0in,width=3.25in}}
\vspace{0.1in}
\caption{The sliding time $t^s$ of the particle $i$.  The solid line shows the
result for a symmetric collision. Here $e_n = 0.9$ and $C=0.13$.}
\end{figure}

An interesting effect can be observed in Fig. 9: there is
a particular combination of the initial particle velocities that gives vanishing sliding
distance.  The meaning of this result is that the particle returns to its initial
position exactly at the time $t^s_i$ after the collision; this occurs when 
${\bf v}_i^s = -{\bf v}_i^f$.  Using Eqs.~(\ref{eq:slip_vel_sa},~\ref{eq:vfin}), we obtain the 
condition for zero sliding distance in terms of the initial velocities of the particles
\be
{\bf v}_i^0 = {(e_n + 2 ) \left(1+{mR^2\over I}\right) +e_n {mR^2\over I} -2 C
\over (e_n - 2 ) \left(1+{mR^2\over I}\right) +e_n {mR^2\over I} +2 C} {\bf v}_j^0
\ .
\label{eq:s0}
\ee
For a completely elastic collision of solid particles, we obtain
${\bf v}_i^0 = -6 {\bf v}_j^0$.  Equation~(\ref{eq:s0}) gives clear experimental
prediction which can be used to explore how realistic the collision model is.

\subsubsection {The change of the translational kinetic energy and momentum due 
to sliding}

In this section we give the final results for the change of the 
translational energy and the linear momentum of the particles due to sliding 
after a collision.
This results assume that the particles slide the whole distance $s$, so 
that there are no other collisions taking place while the particles
travel this distance.  Consequently, for a system consisting of
many particles (as in~\cite{golob,behringer}), the change of the translational energy 
due to sliding depends on the distance traveled by the particles in between of 
the collisions, $l$.  When $l$ is on average much larger than the
sliding distance, $s$, one could consider modeling the effect of sliding using
``effective'' coefficient of restitution~\cite{golob}, which we derive below.  
In this case, we find that this ``effective'' coefficient of restitution 
depends only on the usual restitution coefficient, $e_n$, and on the geometric 
properties of the particles.  On the other hand,
if $l\approx s$, this ``effective'' coefficient of restitution will depend
also on the local density and velocity of the particles.
We explore this effect in more details in~\cite{we}.
\begin{figure}[htb]
\centerline{\psfig{figure=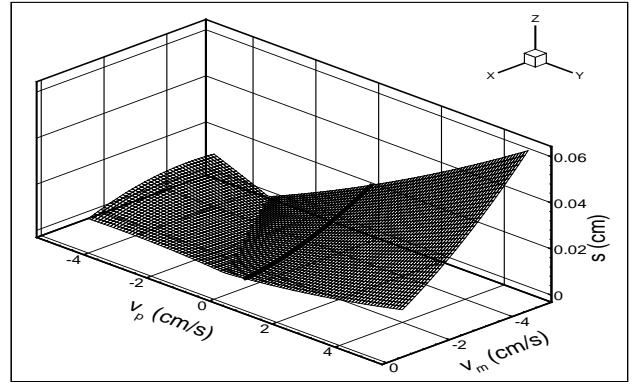,height=2.0in,width=3.25in}}
\vspace{0.1in}
\caption{The sliding distance s of the particle $i$ ($|{\bf s}_i|$ in the text).
The solid line shows the result for a symmetric collision ($e_n = 0.9$ and
$C=0.13$).}  
\end{figure}

The change of the translational energy of the particle $i$, $\Delta\bar E_{slip}^i$, 
is defined as $\Delta\bar E_{slip}^i = |{\bf v}_i^f|^2 - |{\bf v}_i^s|^2$.
The translational
velocity of the particle when it stops sliding, ${\bf v}_i^s$, is given by
Eq.~(\ref{eq:slip_vel_sa}), and the 
velocity of the particle at the end of collision,
${\bf v}_i^f$, is given by Eq.~(\ref{eq:vfin}).  
\begin{figure}[htb]
\centerline{\psfig{figure=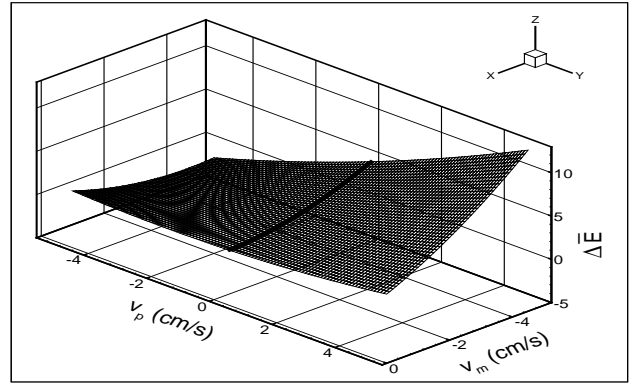,height=2.0in,width=3.25in}}
\vspace{0.1in}
\caption{The change of the translational energy of the particle $i$ due to sliding
($\Delta\bar E_{slip}^i$ in the text).
The solid line shows the result applicable to symmetric collisions  
($e_n = 0.9$ and $C=0.13$).}
\end{figure}

Figure 10 shows the results for $\Delta \bar E_{slip}^i$.  We chose to show the total energy
change, instead of the relative one, in order to be able to address the case of
initially stationary particle, characterized by $E_0^i = 0$.  The solid line shows the result for
the symmetric case, $v_p  = 0$.  From Fig. 10 we observe that the loss of energy of the particle 
strongly depends on $v_p$, i.e. on the degree of the asymmetry of
the collision.  In particular, we observe that $\Delta \bar E_{slip}^i$ could attain
{\it negative} values, meaning that the particle {\it increases its translational kinetic
energy due to sliding}.  In order to illustrate this rather
counter-intuitive point, let us consider for a moment completely asymmetric collision, 
characterized by ${\bf v}_j^0 = v^0{\bf\hat i}$, ${\bf v}_i^0 = 0$.
Using Eqs.~(\ref{eq:slip_vel_sa},~\ref{eq:vfin}), the change of 
the energy of the particle $i$ (initially stationary particle), due to sliding, 
easily follows
\begin{eqnarray}
\Delta \bar E_{slip}^i && =
{{1\over 2} (1+e_n) -C\over \left( 1 + {m R^2\over I}\right)^2}\times \nonumber \\
&& \left[ (1+e_n) \left( {1\over 2} + {m R^2\over I}\right) - C\right] (v^0)^2\ .
\label{eq:slip1}
\end{eqnarray}
Since $C=O(\epsilon)\ll 1$, $\Delta \bar E_{slip}^i$ is
positive, meaning that the particle $i$ loses its translational energy due to
sliding after the collision. On the other
hand, the change of the energy of the particle $j$ (impact particle), due to
sliding, is given by
\begin{eqnarray}
\Delta \bar E_{slip}^j &&= -\,
{ {1\over 2} (1+e_n) -C\over \left( 1 + {m R^2\over I}\right)^2}\times \nonumber \\
&& \left[ (1-e_n)\left( 1 + {m R^2\over I}\right) 
 + {1+e_n\over 2} - C\right] (v^0)^2\ .
\label{eq:slip2}
\end{eqnarray}
The negative sign implies that the particle $j$ {\it gains} translational
energy by sliding.  The interpretation of this result is simple, in particular in
completely elastic limit, $e_n\rightarrow 1$ (also $C\rightarrow 0$). Since the collision
is elastic, the translational velocity of the impact particle $j$ vanishes immediately 
after the collision with the stationary particle $i$.  But, the particle $j$ still has
the angular velocity, ${\bf \Omega}^f_j$, which is (in the elastic limit) equal to
its initial angular velocity.  Consequently, the particle $j$ has the sliding 
velocity, which is, immediately after the collision, equal to the negative of its 
initial translational velocity.  The sliding acceleration resulting from this sliding
velocity induces the motion of the particle in its initial, ${\bf \hat i}$, direction.  
The result is that the translational energy of the particle $j$ is being increased by the 
action of the friction force between the particle and the substrate after the collision.

Still considering completely asymmetric case, it is of interest to compute the net
energy loss of the system of two particles, 
$\Delta \bar E_{slip}^{i,j} = \Delta \bar E_{slip}^i+ \Delta \bar E_{slip}^j$.
By combining Eqs.~(\ref{eq:slip1},~\ref{eq:slip2}), we obtain
\begin{eqnarray}
\left[ \Delta \bar E_{slip}^{i,j}\right]^{asymm} &&=
{ {1\over 2} (1+e_n) -C\over \left( 1 + {m R^2\over I}\right)^2}
\left[ (1+e_n) {m R^2\over I}-\right. \nonumber \\
&& \left. (1-e_n)\left( 1 + {m R^2\over I}\right) \right] (v^0)^2\ .
\end{eqnarray}
The net change of the translational energy is positive, as 
expected, so that the system is losing translational kinetic energy. 
As in the symmetric case, we obtain the relative loss of energy by dividing with
the total initial translational kinetic energy (scaled with reduced mass),
$\Delta E_{slip}^{i,j} = {\Delta \bar E_{slip}^{i,j}/(v^0)^2}$.
In the completely elastic case, the result for the relative loss of energy is
given by
\be
\left[ \Delta E_{slip}^{i,j}\right]_{elastic}^{asymm} =
{2 {m R^2\over I}\over \left( 1 + {m R^2\over I}\right)^2} \ .
\label{eq:e_as}
\ee
Following the same approach, the relative loss of energy of
the system of two particles undergoing a symmetric elastic collision 
(scaled with the total initial energy) is given by 
(using Eqs.~(\ref{eq:v_s},~\ref{eq:eloss_s}))
\be
\left[ \Delta E_{slip}^{i,j}\right]_{elastic}^{symm} =
{4 {m R^2\over I}\over \left( 1 + {m R^2\over I}\right)^2} \ .
\label{eq:e_s}
\ee
Comparing Eqs.~(\ref{eq:e_as},~\ref{eq:e_s}), we see that the particles
lose twice as much energy due to sliding in symmetric, compared to completely
asymmetric elastic collision.  The intuitive understanding of this result
follows by realizing that the sliding velocities of the particles at the
end of a symmetric collision, scaled by the initial velocities, are larger
in the symmetric, compared to the completely asymmetric case 
(viz. Eq.~(\ref{eq:uf_a})).   The consequence is that the
particles that have undergone a symmetric collision slide longer and lose
more translational energy.  When $C\ne 0$, the loss of energy due to sliding in an
inelastic collision is even smaller, since the particle-particle interaction 
during the collision decreases the angular velocities and, consequently, 
the sliding velocities of the particles after the collision.

Figure 11 shows the change of momentum due to sliding, defined as 
$\Delta \bar p_{slip}^i = ({\bf v}_f^i - {\bf v}_s^i)\cdot{\bf \hat i}$, 
so that it measures the change of the translational velocity of 
particle $i$ (in the ${\bf \hat i}$ direction), after the collision.
Clearly, $\Delta \bar p_{slip}^i$ depends very weakly on the degree of the asymmetry.
For completely elastic collisions, $\Delta \bar p_{slip}^i$ depends only on the relative
initial velocity of the particles, and it is given by
\be
\left( \Delta \bar p_{slip}^i\right)_{elastic} = 
- {1\over 1 + {mR^2\over I}} ({\bf v}_i^0 - {\bf v}_j^0)\cdot {\bf \hat i}\nonumber\ .
\ee
\begin{figure}[htb]
\centerline{\psfig{figure=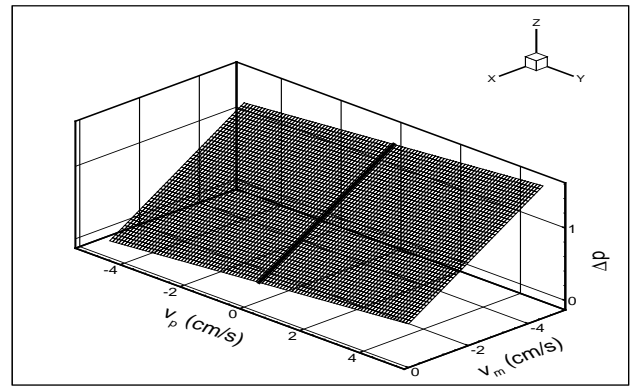,height=2.0in,width=3.25in}}
\vspace{0.1in}
\caption{The change of the linear momentum of the particle $i$ due to sliding.
The solid line shows the result applicable to symmetric collisions  
($e_n = 0.9$ and $C=0.13$).}
\end{figure}

{\it Effective coefficient of restitution}.  Let us define $t_m$ as the time, 
measured from the end of a collision, at which neither of the particles slides anymore, 
so that $t_m= max(t_i^s,\, t_j^s)$, where $t_k^s$ ($k=i,j$) is the sliding
time of a particle, given by Eq.~(\ref{eq:ts_a}). 
We define the effective coefficient of restitution, $e_n^{eff}$, 
as the ratio of the translational velocities of the particles at the time $t_m$, 
and their initial velocities.  Using the translational velocity of the particle $i$, 
given by Eq.~(\ref{eq:slip_vel_sa}), and the analogous equation for the particle $j$,
we obtain
\be
e_n^{eff} = {|{\bf v}_i - {\bf v}_j|^s\over |{\bf v}_i - {\bf v}_j|^0} =
{e_n {mR^2\over I} - 1 \over 1 + {mR^2\over I}}\ .
\label{eq:eff}
\ee
Remarkably enough, this result involves only the ``real'' coefficient of 
restitution and the geometric properties of the particles. 
For solid spheres, the difference between the usual coefficient of restitution and 
the effective one is huge; for  $e_n=0.9$, we obtain 
$e_n^{eff} = 0.36$.   This value is smaller than the range reported in \cite{golob},
but very close to our experimental results for steel particles on aluminum substrate~\cite{ben}.
Slight imperfections from the spherical shape in experiments, non-central collisions, 
and/or the fact that the static friction between the particles has been neglected in our 
calculations, might be the reason for this discrepancy.

{\it General remarks.}
While more precise analysis and material parameters could be used 
in order to more precisely model experiments, we consider that the main 
results and observations given in this section are model-independent.  
In particular, the observation that the sliding is likely to occur as
a consequence of most of the collisions does not depend
on the details of the model.  Of course, the results would be
modified in the case of more complicated (two dimensional) geometry of the 
collisions.  Still, the particular geometry of a collision enters into our 
results for the energy and momentum change only through the observation 
that the frictional interaction of the colliding particles with the 
substrate can be ignored during a collision. 
Since, for the system that we consider in this work, the collision 
forces are generally much stronger than the friction forces resulting 
from particle-substrate interaction, we do not expect this observation 
to be modified for more complicated collisions.  We do note that more 
realistic model for the particle-particle interactions (e.g. by including
static friction) would introduce modifications in the expression for the final
angular velocity of the particles, Eq.~(\ref{eq:omega}).

In the experiments~\cite{golob,behringer} it is observed that
some of the particles travel for long distances without colliding.  Especially in this
situation, it is important to include the effect of rolling friction, which
we have ignored in this section.  As long as a particle slides, the effect
of rolling friction could be safely neglected, since the coefficient of
rolling friction is much smaller than the coefficient of kinematic sliding
friction.  

\section{Conclusion}

The most important observation made in this work is that sliding
leads to a considerable modification of the translational kinetic energy and 
linear momentum of the particles, even in the limit of completely 
elastic collisions.  Based on this observation we give the result
for the ``effective'' coefficient of restitution, valid for
dilute systems, where the mean free path of the particles in
between of the collisions is much longer than the sliding distance.
For more dense systems, we conjecture that this ``effective'' 
coefficient of restitution strongly depends on  
the local density and velocity of the particles.
 
The model that we present is to be used
in molecular dynamics (MD) type simulations~\cite{we} of externally
driven system of a set of particles interacting on a horizontally
oscillated surface.   In particular, we have prepared the grounds
for detailed modeling of the system of two kinds of particles,
which are characterized by different rolling properties.
In~\cite{behringer} is shown that strong segregation
can be achieved.  Preliminary MD results, based on the model
formulated in this paper, show that the realistic modeling
of the particle-particle and particle-substrate interactions
are needed in order to fully understand this effect. 

Further, since experiment is the ultimate test for every theory, it would
be of considerable importance to extend the previous
work~\cite{mcnamara2,mcnamara3,goldhirsch1,ben-naim} on formulating
continuum theory for ``2D granular gas''.  Using the model presented
here should allow for precise comparison between experimental and 
theoretical results.  Possible formulation of realistic 
continuum, hydrodynamic theory applicable to this seemingly simple
system would be an important step towards better understanding of
granular materials.

{\bf Acknowledgments}.  The author would like to thank Robert Behringer,
David Schaeffer, and in particular Joshua Socolar for useful
discussions and comments.  This work was supported by
NSF Grants No. DMR-9321792 and DMS95-04577, and by
ONR Grant No. N00014-96-1-0656.

\appendix

\section{Linear model for the normal force between particles}
\label{sec:linear}

Let us analyze a simple situation, a central collision of two
identical particles, $i$ and $j$, moving with the velocities, 
${\bf v}_i^0$ and ${\bf v}_j^0$, in the ${\bf \hat i}$ direction only.  
Here we ignore the interaction of the particles with the substrate; the importance 
of this interaction is discussed in Appendix~\ref{sec:substrate}.
Using this assumption, the normal force, given by Eq.~(\ref{eq:fn}), is
the only force acting on the particle $i$ in the normal direction.  
By combining the equations of motion for the particles $i$ and $j$, we obtain that 
the compression depth $x = {(d - r_{i,j})/d}$ satisfies the following equation
\be
\ddot x + \gamma_N \dot x + \omega_0^2 x = 0\ ,
\label{eq:ho}
\ee
where $\gamma_N$ is the damping coefficient in the normal direction, and
$\omega_0 = \sqrt {2k/m}$.  We limit our discussion to the case of
low damping, so that $\epsilon = {\gamma_N/\omega_0}\ll 1$.  

This equation is subject to the following initial conditions: 
$x(t=0)=0$, $\dot x (t=0) = {v^0_{rel}/d}$.  The relative velocity of the
particles at $t=0$ is given by $v^0_{rel} = |{\bf v}_i^0 - {\bf v}_j^0|$; 
for a symmetric collision, $v^0_{rel} = 2 v^0$.   The solution is
\be
x = {v^0_{rel} \over d \sqrt{\omega_0^2 - \left ({\gamma_N\over 2}\right)^2}}
\exp \left(-{\gamma_N\over 2} t \right) 
\sin \left( \sqrt{\omega_0^2 - \left ({\gamma_N\over 2}\right)^2} t\right)\ .
\label{eq:comp_depth}
\ee
The duration of the collision, $t_{col}$, now follows from the requirement
$x(t=t_{col}) = 0$, thus
\be
t_{col}= {\pi\over \sqrt{\omega_0^2 - \left ({\gamma_N\over 2}\right)^2}}
= {\pi\over \omega_0} (1 + O(\epsilon^2))\ .
\label{eq:col}
\ee
In what follows, we also need the time of maximum compression, $t_{max}$.
From the condition $\dot x (t=t_{max}) = 0$, we obtain
\be
\tan \left( \sqrt{\omega_0^2 - \left ({\gamma_N\over 2}\right)^2} t_{max}\right) =
{\sqrt{\omega_0^2 - \left ({\gamma_N\over 2}\right)^2}\over {\gamma_N\over 2}}\nonumber \ .
\ee
Expanding to $O(\epsilon)$, it follows
\be
t_{max} = {t_{col}\over 2} \left( 1 - {\epsilon\over \pi} + O(\epsilon^2) \right)\ .
\label{eq:t_max}
\ee
So, the damping manifests itself in a slight asymmetry of the collision, since
$t_{max} < {t_{col}/ 2}$.  The maximum compression,
$x_{max} = x(t=t_{max})$, follows using Eq.~(\ref{eq:comp_depth}).  It is given 
by Eq.~(\ref{eq:max_comp}) for inelastic
collisions, and by Eq.~(\ref{eq:xmax_0}) for elastic ones.

We define the coefficient of restitution as the ratio of the final velocities
of the particles relative to their initial velocities, i.e.
$e_n = {|{\bf v}_i - {\bf v}_j|^f/ |{\bf v}_i - {\bf v}_j|^0}$. It 
follows
\be
e_n = -\, { d\over v^0_{rel}} \dot x(t_{col}) = \exp \left( 
{-{\gamma_N\over 2} t_{col}}\right)\ = 1 - {\pi\over 2}\epsilon + 
O(\epsilon^2)\ .
\label{eq:rest}
\ee
In the limit of low damping, $e_n$ is close to $1$; typically we use $e_n = 0.9$, 
appropriate for steel particles~\cite{golob,behringer,ben}. 
Using Eq.~(\ref{eq:col}), we obtain 
$\epsilon \approx - {2/\pi} \ln (e_n) \approx 0.07$.

The final velocity of the particle $i$ (at the end of the collision) follows
from the requirement that the total linear momentum is conserved in the center
of mass frame.  It is given by 
\be
{\bf v}_i^f = {1\over 2} \left[ {\bf v}_i^0 + {\bf v}_j^0
- e_n ({\bf v}_i^0 - {\bf v}_j^0)\right]\ .
\label{eq:vfin}
\ee
For a symmetric collision,  this results simplifies to 
${\bf v}_i^f = - e_n {\bf v}_i^0$.

\section{Nonlinear models for the normal force between particles}
\label{sec:nonlinear}

The linear model, presented in the previous section, is the simplest
approximation for the collision interaction between particles.  
Nonlinear terms, resulting from the final area of contact and other
effects~\cite{luding,ristow,taguchi,brilliantov1,poschel}, 
should be included in order to model the
interaction between particles more realistically.  Still, since we
are concerned with the conditions where the maximum compression depth is small,
the linear approximation is a reasonable one.  We use the nonlinear model,
outlined below, in order to connect the values of the parameters, in particular
the collision time, $t_{col}$, with the material properties of the particles.

The general, commonly used equation is~\cite{luding}
\be
\ddot x + {\eta d\over m}x^\gamma 
\dot x + {E d\over m} x^{\beta+1} = 0\nonumber \ ,
\ee
where $\eta$ and $E$ are the material constants. The
choice $\gamma=0$, $\beta=0.5$ leads to the Hertz model.
The analysis of this equation gives an
expression for $t_{col}$, that can then be used to determine
the appropriate force constant in the linear model, $k$, and the damping
coefficient, $\gamma_N$.  The result for the collision time is~\cite{luding}
\be
t_{col}\approx I(\beta) \left( 1+{\beta\over 2}\right)^{1\over 2 + \beta}
\left( {m\over E d^{1-\beta}}\right)^{1\over 2 + \beta}
v_0^{-{\beta\over 2 + \beta}}\nonumber \ .
\ee
For the Hertz model, $I(0.5)=2.94$.  The parameter 
$E$ is given by ${Y/ (3(1-\tilde\sigma^2))}$, where $Y$ is the Young modulus, and
$\tilde\sigma$ is the Poisson ratio.  We use $Y=2.06\times 10^{12}$ dyn/cm$^2$,
and $\tilde\sigma = 0.28$.  For steel spheres with diameter $d = 4\, $mm, and impact 
velocity, $v_0 =10\, $ cm/s, $t_{col}\approx  2.55\times 10^{-5}\, $ sec; for
$v_0 =100\, $ cm/s, $t_{col}\approx 1.61\times 10^{-5}\, $ sec.
We note that the model predicts $t_{col} \sim v_0^{-1/5}$, and $t_{col}\sim d$.
The parameters that enter the linear model can now be calculated, using
$\omega_0 = {\pi/t_{col}}(1+ O(\epsilon^2))$, and 
$\gamma_N= -{2/t_{col}} \ln (e_n)$. 

\section{Sliding during collisions}
\label{sec:approx}

\subsection{Sliding during a symmetric collision}

In Appendices~\ref{sec:linear} and~\ref{sec:nonlinear} we obtained the results
governing dynamics of particle collisions, ignoring the interaction with 
the substrate.  Here we show that the colliding particles slide through most of a 
typical collision. The additional material constants that are involved 
are the coefficients of static and kinematic friction between the
considered particles and the substrate, $\mu_s$ and $\mu_k$. 
In our estimates, we use $\mu_s = 0.5$ and $\mu_k = 0.1$.

The condition for sliding, Eq.~(\ref{eq:slip_res}), applied to the simple
situation outlined in section~\ref{sec:symmetric}, gives that sliding
occurs when $|{\bf F}_N^c| \ge (1 + {m R^2/ I})\, \mu_s m g$.  
In terms of the compression depth and velocity, this condition is
\be
d\omega_0^2 x + d \gamma_N \dot x \ge \left( 1 + {m R^2\over I}\right) \mu_s g\ .
\label{eq:slipx}
\ee
We note that the left hand side of this equation is always non-negative, since
${\bf F}_N^c$ is always repulsive (at the very end of a collision, when
$x\ll 1$, $\dot x < 0$, ${\bf F}_N^c$ is set to $0$).
In the limit $\gamma_N \rightarrow 0$, we obtain that sliding occurs
when $x \ge x_{min}^0$, where $x_{min}^0 = (1 + {m R^2/I}) {\mu_s g/(d \omega_0^2)}$.  
Using the result for the compression depth, Eq.~(\ref{eq:comp_depth}),
we obtain the time at which sliding starts, $t_{min}^0$, measured from the beginning
of the collision, 
\be
\sin (\omega_0 t_{min}^0) = \left( 1 + {m R^2\over I}\right ) 
{\mu_s g\over v^0_{rel} \omega_0}\nonumber \ .  
\ee
For the initial velocities, $v^0_{rel}$, satisfying $v^0_{rel}\gg v^b$, where
$v^b = (1 + {m R^2/ I}) {\mu_s g/ \omega_0}$, it follows that 
$\sin (\omega_0 t_{min}^0) \ll 1$.
For our set of parameters, and assuming solid spheres, 
$v^b\approx 10^{-2}$ cm/s.  Therefore, this condition
is satisfied for most of the collisions. 
Assuming this, we obtain
\be
t_{min}^0= \left (1 + {m R^2\over I}\right)  
{ \mu_s g\over v^0_{rel} \omega_0^2}\ ,
\label{eq:t_min0_0}
\ee
and
\be
x_{min}^0 = \left ( 1+{mR^2\over I}\right ) 
{\mu_s g\over d \omega_0^2} \ .
\label{eq:xmin0_ap}
\ee
Exploiting the symmetry of an elastic collision, we conclude that the sliding
condition is satisfied for $t_{min}^0 < t < t_{col}-t_{min}^0$.

Next we go to the limit of small, but finite damping, and assume that
the condition $\omega_0 t_{min}\ll 1$ is still valid, where $t_{min}$ is
now the time when sliding occurs for $\gamma_N \ne 0$.  Using
$\gamma_N t_{min} \ll \omega_0 t_{min}$, we Taylor-expand
$x$ and $\dot x$ (given by Eq.~(\ref{eq:comp_depth})) 
at $t=t_{min}$, and keep 
only the first order terms in small quantities $\omega_0 t_{min}$,
$\gamma_N t_{min}$.  In this limit,
\be
x(t_{min})\approx {v^0_{rel}\over d} t_{min};\quad 
\dot x(t_{min}) \approx {v^0_{rel}\over d} (1 -\gamma_N t_{min})\ .
\label{eq:xapprox}
\ee
The sliding condition, 
Eq.~(\ref{eq:slipx}), gives the time when sliding 
occurs, for an inelastic collision
\be
t_{min}= \left (1 + {m R^2\over I}\right)  
{ \mu_s g\over v^0_{rel} \omega_0^2} - {\epsilon\over \omega_0}
+ O(\epsilon^2)\ .
\label{eq:tmina}
\ee
We note that there are two factors that contribute to $t_{min}$: the frictional
interaction with the substrate gives the first term on the right hand side of
Eq.~(\ref{eq:tmina}), and the damping that occurs during a collision gives
the second one.   For the initial velocities, satisfying
$v^0_{rel} \gg v^c ={g\mu_s/\gamma_N}$, the contribution from the damping is the
important one.  Using the expression for $\gamma_N$ given
in Appendix~\ref{sec:nonlinear}, we obtain $v^c\approx 0.05$ cm/s (for
$e_n=0.9$).  This velocity is smaller than the usual initial velocities
considered in this work.  Assuming $v^0_{rel} \gg v^c$, we conclude that the
friction term could be relevant only in the limit $e_n \rightarrow 1$, since $v^c$ diverges in 
this limit. Consequently, it follows that $t_{min} \rightarrow 0$, so that the sliding starts 
immediately at the beginning of an inelastic symmetric collision.  Since
$t_{min} \rightarrow 0$, the expansion used to obtain Eq.~(\ref{eq:xapprox}) is consistent.

\subsection{Sliding during an asymmetric collision}

By combining Eqs.~(\ref{eq:e.00a},~\ref{eq:fn_a}), we obtain
the condition for sliding during an asymmetric collision
\be
{\left| {\bf F}_{N,i}^c - {m R^2\over I} ({\bf \hat k}\cdot {\bf F}_{
R,i}^c){\bf \hat n}\right|
\over  1 + {mR^2\over I}} \ge
\mu_s \left| mg  - {\bf F}_{R,i}^c\cdot {\bf \hat k}\right |\nonumber \ .
\ee
Using 
Eqs.~(\ref{eq:fn},~\ref{eq:force_r1}) for ${\bf F}_{N,i}^c$ and 
${\bf F}_{R,i}^c$, respectively, we obtain (in terms of the compression
depth, see Appendix~\ref{sec:linear})
\begin{eqnarray}
d\omega_0^2 x + d\gamma_N \dot x \ge &&
\left(  1 + {mR^2\over I}\right) \mu_s g +
{\gamma_S\over 2} |v_{rel}^z| sign(-v_{rel}^z)\times\nonumber \\ 
&&\left[ {mR^2\over I} - \left(1 + {mR^2\over I}\right)\mu_s\right]\ ,
\label{eq:cond}
\end{eqnarray}
where $v_{rel}^z$ is given by Eq.~(\ref{eq:vrelz}).  From the first part of this 
Appendix, we already know that the first term on the right hand side is
negligible. The term inside the square brackets is positive for solid spheres, and 
$ \mu_s = 0.5 $. For large $x$'s, the condition, Eq.~(\ref{eq:cond}), is always satisfied, 
since $d\omega_0^2 x$ is the dominant term.  So, we need to explore only the beginning and end of
a collision.  If $sign(-v_{rel}^z)< 0$, the sliding condition is always satisfied; 
so that the slower particle always slides.  When $sign(-v_{rel}^z) > 0$, we concentrate on 
the very beginning of the collision, and obtain the condition
\be
\gamma_N \ge {\gamma_S\over 2} {|{\bf v}_i^0 + {\bf v}_j^0|\over
|{\bf v}_i^0 - {\bf v}_j^0|} 
\left[ {mR^2\over I} - \left(1 + {mR^2\over I}\right)\mu_s\right]\nonumber \ .
\ee
Since typically $\gamma_S = {\gamma_N/2}$, this condition is
satisfied, assuming $|{\bf v}_i^0 + {\bf v}_j^0|\approx |{\bf v}_i^0 - {\bf v}_j^0|$.  

We conclude that the particles entering an asymmetric collision slide during
the whole course of the collision, except possibly in the case 
$|{\bf v}_i^0 - {\bf v}_j^0|\ll |{\bf v}_i^0 + {\bf v}_j^0|$. We do not consider
this case here.

\section{Modification of collision dynamics due to the interaction with
the substrate}
\label{sec:substrate}

Here we estimate the importance of the interaction between the colliding 
particles and the substrate during a collision.  In particular, we estimate
under what conditions the interaction with the substrate significantly modifies
the results for the compression depth and the duration of a collision.
We use the linear model outlined in Appendix~\ref{sec:linear}, and concentrate on
the case of the particles moving on a horizontal static substrate.  

In Appendix~\ref{sec:approx} it is shown that, assuming typical 
experimental conditions, the colliding particles slide relative to the substrate during 
most of a collision.  For simplicity, here we concentrate on a symmetric 
collision, and further assume that the condition for sliding is satisfied throughout the 
collision, so that the friction force attains its maximum allowed value, given by 
Eq.~(\ref{eq:force_slip1}).  By using this approximation, we slightly overestimate
the influence of the friction with the substrate on the dynamics of a collision.

From Fig. 4 we observe that the friction force, $\bf f$, acts in the 
direction opposite to the normal collision force, ${\bf F}_N^c$.  Including
$\bf f$ in the Newton equations of motion for the particles $i$ and $j$, we
obtain the modified equation for the compression depth
\be
\ddot x + \gamma_N \dot x + \omega_0^2 x -\mu_k{g\over d} = 0\ ,
\label{eq:ho1}
\ee
which simplifies to Eq.~(\ref{eq:ho}) if the particle-substrate interaction is
ignored.

Using the initial conditions as in Appendix~\ref{sec:linear}, we obtain the
solution
\begin{eqnarray}
&& x = x_f - \exp \left(-{\gamma_N\over 2} t \right)
\left[ x_f
\cos\left( \sqrt{\omega_0^2 - \left ({\gamma_N\over 2}\right)^2} t\right) - \right.
\nonumber \\ && \left. {{v^0_{rel}\over d} - {1\over 2} x_f \gamma_N\over 
\sqrt{\omega_0^2 - \left ({\gamma_N\over 2}\right)^2} }
\sin \left( \sqrt{\omega_0^2 - \left ({\gamma_N\over 2}\right)^2} t\right)\right]\ ,
\label{eq:comp_depth1}
\end{eqnarray}
where $x_f = {\mu_k g /(d\omega_0^2})$.  

{\it Collision time.}  For simplicity, we concentrate on the case of
zero damping ($\gamma_N = 0$), and calculate the change of the duration of the
collision due to the particle-substrate interaction.  Let us assume that the change
of the collision time is small, and write $t_{col}' = t_{col} + \tau$, where
$t_{col}={\pi/\omega_0}$ is the collision time if there is no interaction with
the substrate, and $\tau \ll t_{col}$.  Using the condition
$x(t = t_{col}') = 0$, and expanding the compression depth, given by
Eq.~(\ref{eq:comp_depth1}), to the first order in the small quantity  
$\tau \omega_0$, we obtain that $\tau = {2 x_f d /v^0_{rel}}$.  
So, the relative change of the collision time due to the interaction
with the substrate is given by
\be
{t_{col}' - t_{col}\over t_{col}} = {2 \mu_k g \over \pi v^0_{rel} 
\omega_0}\nonumber \ .
\ee
For $v^0_{rel}\gg v^a$, where 
$v^a \approx {\mu_k g / \omega_0}$, the change of 
the collision time is small.  Using the parameters given in 
Appendices~\ref{sec:nonlinear} and \ref{sec:approx}, we estimate
$v^a\approx 10^{-3}$ cm/s.  So, for most of the experimentally realizable
conditions, the duration of a collision is just very weakly influenced by the
particle-substrate interaction.  We assume $v^0_{rel}\gg v^a$, so that
$\tau \omega_0 \ll 1$, and the expansion of Eq.~(\ref{eq:comp_depth1}) is
consistent.

{\it Maximum compression depth.} Following the same approach, 
we estimate the modification of the maximum compression achieved during
a collision, due to the interaction with the substrate.  Working in the
limit of zero damping, and assuming a small modification of the time, $t_{max}$, when
the maximum compression, $x_{max}'$, is reached, we obtain
$x_{max}' \approx x_f + {v^0_{rel}/(d\omega_0)}$.
Comparing this result with the result for the compression depth calculated
previously, given by the elastic limit of Eq.~(\ref{eq:max_comp}), we obtain
\be
{x_{max}' - x_{max}\over x_{max}} = d \omega_0
{x_f \over v^0_{rel}} = {\mu_k g \over v^0_{rel} \omega_0}\nonumber \ .   
\ee
Similarly to the analysis of the collision time, we observe that 
for $v^0_{rel} \gg v^a$, the maximum compression depth is very weakly
influenced by the particle-substrate interaction.

We conclude that for most of collisions occurring in experiments, 
the interaction with the substrate just slightly modifies the compression
depth and the duration of a collision.  These small
modifications are ignored in the subsequent analysis. 

\section{Rotations of the particles during a collision}
\label{sec:rotations}

\subsection{Rotations during symmetric collisions}

During symmetric collisions, the rotational motion of the particles is influenced
only by the friction force between the particles and the substrate.  
Here we consider only elastic collisions, since in Appendix~\ref{sec:approx} it is
shown that the particles entering an inelastic collisions start sliding
immediately, so that the angular acceleration is constant during the whole
course of collision, simplifying the calculations (see section~\ref{sec:symmetric_inelastic}).
Since there is no possibility of confusion, we use scalar notation, with the 
sign convention that +sign corresponds to the forces acting in 
$+{\bf \hat i}$ direction, and to the angular motion 
in $+{\bf \hat j}$ direction (the coordinate axes are as shown in Fig. 4). 

At the very beginning of an elastic collision, 
for $0<t< t_{min}^0$, ($t_{min}^0$ is given by Eq.~(\ref{eq:t_min0_0})), 
the colliding particles do not slide.  During this time interval,
the angular acceleration of the particle $i$, which initially moves in
$-{\bf \hat i}$ direction, is given by
\be
\dot \Omega_i =  {R\over I} f_i = {R\over I}{F_{N,\, i}^c\over 1+ {mR^2\over I}} =
{{mR\over I}\over 1+ {mR^2\over I}}
v^0_{rel} \omega_0^2 t\nonumber \ ,
\ee
where $x\approx t {v^0_{rel}/d}$, and  Eqs.~(\ref{eq:fn},~\ref{eq:e.00}) have been used.
Integration yields
\be
\Omega_i (t = t_{min}^0) = \Omega_i^0 + {1\over 2} {mR\over I}
\left( 1+ {mR^2\over I}\right) {(\mu_s g)^2\over v^0_{rel} \omega_0^2} \ ,
\label{eq:omega1}
\ee
and $\Omega_i^0 = -{v^0/R}$.  For $t_{min}^0<t<t_{col}-t_{min}^0$, the sliding
condition, $|{\bf f}_i| = \mu_s |{\bf F}_{N,\, i}^c|$, is satisfied, so that
the angular acceleration reaches its maximum (constant) value
\be
\dot \Omega_i = {mR\over I}\mu_k g\ . 
\label{eq:alpha_a}
\ee
For $t_{col}>t>t_{col}-t_{min}^0$, the sliding condition is not satisfied anymore, but
the particle is sliding already, so that $\dot \Omega_i$ is still given by 
Eq.~(\ref{eq:alpha_a}).
The angular velocity of the particle $i$ at the end of collision is
\be
\Omega_i^{f0} = \Omega_i (t = t_{min}^0) + {mR\over I} \mu_k g (t_{col}-t_{min}^0) \ .
\label{eq:omega2}
\ee
Combining Eqs.~(\ref{eq:omega1},~\ref{eq:omega2}), we obtain the final result, given
by Eq.~(\ref{eq:omega_0}).

\subsection{Rotations during asymmetric collisions}

\subsubsection{About tangential force }

Here we estimate under what conditions, ${\bf F}_R^c$, given by Eq.~(\ref{eq:force_r}),
reaches its maximum allowed value, $\nu_k |{\bf F}_N^c| $.  
As mentioned in section~\ref{sec:assymetric}, here we ignore the frictional interaction
of the particles with the substrate during a collision.  For simplicity,
we also neglect the damping in the normal directions, so that $|{\bf F}_N^c| = md \omega_0^2 x$
(see Appendix~\ref{sec:linear}).  Next, we note that the relative velocity of the 
point of contact satisfies $v_{rel}^z (t=0) > v_{rel}^z (t>0)$, since ${\bf F}_R^c$ always 
decreases  $v_{rel}^z$ (given by Eq.~(\ref{eq:vrelz})). 
In what follows, we use $v_{rel}^z (t>0)=v_{rel}^z (t=0)$, and give the upper limit of
the first term entering the definition of ${\bf F}_R^c$.

Let us first concentrate on large compression
depths, $x\approx x_{max} = {|{\bf v}_i^0 - {\bf v}_j^0|/(d\omega_0)}$ 
(see Appendix~\ref{sec:linear}).  This compression is reached at $t=t_{max}={\pi/(2\omega_0)}$.  
We use  $v_{rel}^z (t=t_{max})= v_{rel}^z (t=0)=
|{\bf v}_i^0 + {\bf v}_j^0|$, and obtain that  ${\bf F}_R^c$ reaches its 
maximum allowed value if (see Eq.~(\ref{eq:force_r}))
\be
{\gamma_S\over 2} |{\bf v}_i^0 + {\bf v}_j^0| \ge \nu_k \omega_0 |{\bf v}_i^0 - 
{\bf v}_j^0| \ .
\ee
Since ${\gamma_S/\omega_0}\ll 1$, this condition is never satisfied  for 
$\nu_k = O(1)$, and
$|{\bf v}_i^0 + {\bf v}_j^0|\approx |{\bf v}_i^0 - {\bf v}_j^0|$.  

For small $x$'s, let us assume again $v_{rel}^z (t>0)= v_{rel}^z (t=0)$.
From  Eq.~(\ref{eq:force_r}) it follows that ${\bf F}_R^c$ reaches its cutoff 
value when $x<x^{crit}$, where
\be
{x^{crit}\over x_{max}} = {\gamma_S\over 2\nu_k\omega_0} 
{|{\bf v}_i^0 + {\bf v}_j^0|\over
|{\bf v}_i^0 - {\bf v}_j^0|} = O(\epsilon) \nonumber \ .
\ee
Using $x\approx {t |{\bf v}_i^0 - {\bf v}_j^0|/d}$ (valid for $x\ll x_{max}$), we obtain
that the condition $x<x^{crit}$ is satisfied for $t<t^{crit}$, where 
\be
{t^{crit}\over t_{max}} = { \gamma_S\over\pi\omega_0} {|{\bf v}_i^0 + {\bf v}_j^0|\over
|{\bf v}_i^0 - {\bf v}_j^0|} = O(\epsilon)\nonumber \ . 
\ee
In order to calculate the angular velocity of the particle $i$ at the end of a
collision, ${\bf \Omega}_i^f$, we have to integrate the angular 
acceleration, ${\bf \dot \Omega}_i$, during the course of a collision. 
The angular acceleration is proportional to ${\bf F}_R^c$, as it follows from
Eq.~(\ref{eq:ang_accs1_a}), where ${\bf f}_i$ is being neglected.  
In performing the integration, it appears that we have to consider
separately two regions: $0<t<t^{crit}$, during which ${\bf F}_R^c$ varies, and
$t>t^{crit}$, during which ${\bf F}_R^c$ is constant. 
The final angular velocity of the particle $i$ is formally given by 
\be
{\bf \Omega}_i^f = {\bf \Omega}_i^0 +\int_0^{t^{crit}}{\bf \dot \Omega}_i dt
+ \int_{t^{crit}}^{t_{col}} {\bf \dot \Omega}_i dt \ .
\label{eq:int}
\ee
This result can be simplified by realizing that $|{\bf F}_R^c|=O(\epsilon)$.  
It follows that $|{\bf \dot \Omega}_i| = O(\epsilon)$, 
so that the contribution of the second term on the right
hand side of Eq.~(\ref{eq:int}) is proportional to
$|{\bf \dot \Omega}_i| t^{crit} = O(\epsilon^2)$. 
For consistency reasons we neglect this correction, and ignore 
the fact that $|{\bf F}_R^c|$ could reach Coulomb cutoff at the
very beginning and end of a collision.  This estimate is not valid for 
$|{\bf v}_i^0 - {\bf v}_j^0|\ll |{\bf v}_i^0 + {\bf v}_j^0|$,
so when the particles initially move with almost the same velocities.  
As already mentioned in Appendix~\ref{sec:approx}, we do not consider this case here.

\subsubsection{The angular velocity of the particles during an asymmetric collision}

Using Eqs.~(\ref{eq:ang_accs1_a},~\ref{eq:force_r1}), and
neglecting the particle-substrate interaction during a collision,
we obtain the angular acceleration of the particle $i$,  
\be
{\bf \dot \Omega}_i = -\, {mR^2\over I} {\gamma_S\over 2}
[({\bf \Omega}_i + {\bf \Omega}_j)\cdot {\bf \hat j}]\ {\bf \hat j} \nonumber \ ,
\ee
and ${\bf \dot \Omega}_j = {\bf \dot \Omega}_i$.  Recalling that ${\bf \dot \Omega}_i$ and
${\bf \dot \Omega}_j$ are always in the opposite direction from 
${\bf \Omega}_i + {\bf \Omega}_j$, we obtain a simple system of 
coupled ordinary differential equations 
\begin{eqnarray}
&&{\bf \dot \Omega}_i = -\, C' ({\bf \Omega}_i + {\bf \Omega}_j)\ ,\\
&&{\bf \dot \Omega}_j = -\, C' ({\bf \Omega}_i + {\bf \Omega}_j)\ ,
\end{eqnarray}
where $C' = {mR^2\gamma_S/(2I)}$.  We define ${\bf \Omega}_+={\bf \Omega}_i +
{\bf \Omega}_j$, so that 
${\bf \dot \Omega}_+ = -\, 2 C' {\bf \Omega}_+$, 
with the solution \be
{\bf \Omega}_+(t) = {\bf \Omega}_+(t=0) \exp(-2 C' t)\nonumber \ .
\ee  
At $t=t_{col}$, ${\bf \Omega}_+(t=t_{col})={\bf \Omega}_+(t=0) \exp(-2 C)$, where
$C=C'\, t_{col} = O(\epsilon)$.
Recalling that the changes of ${\bf \Omega}_i$ and ${\bf \Omega}_j$ are the
same, so that ${\bf \Omega}_k(t=t_{col}) = {\bf \Omega}_k(t=0) +\Delta{\bf \Omega}$, 
($k=i,j$), the change of the angular velocities is given by
\be
\Delta {\bf \Omega} = {1\over 2}\left[ ({\bf \Omega}_i^0 + {\bf\Omega}_j^0)
(\exp (-2C) -1)\right] \approx -({\bf \Omega}_i^0 + {\bf \Omega}_j^0) C \nonumber \ ,
\ee
correct to the first order in $\epsilon$.  For $\gamma_S={\gamma_N/2}$,
and the parameters as in Appendix~\ref{sec:nonlinear}, 
$C={\pi m R^2 \epsilon/(4 I)}\approx 0.13$.
The final angular velocity of the particle $i$ is now given
by Eq.~(\ref{eq:omega}). 

\section{Jump condition for asymmetric particle collisions}
\label{sec:jump}

Throughout this work, we have assumed that the particles are bound to move 
on the surface of the substrate.  Here we explore the validity of this 
assumption.
The required condition for a particle to be bound to the substrate is that
the normal force $|{\bf F}_N|$, given by Eq.~(\ref{eq:fn_a}), is nonzero.    
We immediately observe that only a particle colliding with a slower 
particle (so that $sign(-v_{rel}^z)<0$, see Eq.~(\ref{eq:vrelz})) 
experiences a force in $+{\bf \hat k}$ direction, due to a collision.  
Let us concentrate on this situation. Using the value
of $|{\bf F}_R^c|$ at $t=0$, we obtain that
a particle detaches from the substrate if
\be
f_{net} = m \left( {\gamma_S\over 2} |v_{rel}^z| - g\right)\nonumber > 0 \ ,
\ee
where Eqs.~(\ref{eq:vrelz},~\ref{eq:fn_a},~\ref{eq:force_r1}) have been used.
It follows that, during the collisions distinguished by
$|v_{rel}^z| > v^d = {2g/\gamma_S}$, the faster particle 
detaches from the substrate.  Using the values of the parameters as in
Appendix~\ref{sec:nonlinear}, and $\gamma_S = {\gamma_N/2}$, 
we obtain $v^d\approx 0.5 $ cm/s.  Correspondingly, this effect takes
place during most of the asymmetric collisions occurring in typical 
experiments~\cite{golob,behringer}.  By relating the impulse of the
force $f_{net}$ transferred to a particle while the collision is 
taking place, with the change of the momentum of the particle in the
${\bf \hat k}$ direction, we obtain the estimate for the initial velocity
of the particle in the ${\bf \hat k}$ direction
\be
v^z = \left( {\gamma_S\over 2} |v_{rel}^z| -g \right) {\pi\over 
\omega_0}\nonumber \ .
\ee
The maximum height above the substrate which the particle reaches is
$h^z = {1/2} {(v^z)^2/g}$, and the time spent without contact with the
substrate is $t^z = {2 v^z/g}$. Let us assume a completely asymmetric collision,
so that $|{\bf v}_i^0| = v^0$, $|{\bf v}_j^0| = 0$, and 
$v_{rel} = v_{rel}^z = v^0$.  Using the parameters from 
Appendix~\ref{sec:nonlinear}, for $v^0 = 10$ cm/s, we obtain
$v^z\approx 0.5$ cm/s, $h^z\approx 1.3\times 10^{-4}$ cm, and
$t^z \approx 10^{-2}$ sec.  Since the maximum height is much 
smaller than the diameter of the particles, this detachment 
introduces negligible corrections to the dynamics of the particle
collisions in $x-y$ plane.  Further, even though $t^z\gg t_{col}$, so that the
particle is not in the contact with the substrate during the time which
is much longer than the duration of the collision, $t^z$ is still much smaller
then the sliding time scale, specified by Eq.~(\ref{eq:ts_a}).
So, our results for sliding of the particles after a collision 
are not significantly modified due to the detachment effect.


\end{document}